\documentclass[preprint,12pt]{aastex}

\newif\ifpp\pptrue

\renewcommand{\deg}{\ifmmode^\circ\else$^\circ$\fi}
\newcommand{\et}{et al.}

\newcommand{\um}{\hbox{$\mu$m}}

\input{psfig.sty}
\slugcomment{ApJ Press Version, Mar 2001}

\begin{document}

\title{Interpreting the Cosmic Infrared Background: Constraints on the Evolution of the
Dust Enshrouded Star Formation Rate} 

\author{R.  Chary} 
\affil{Department of Astronomy \& Astrophysics, University of California, Santa Cruz, CA 95064}

\and

\author{D. Elbaz\altaffilmark{1,}\altaffilmark{2}} 
\affil{CEA Saclay - DAPNIA - Service
d'Astrophysique, Orme des Merisiers, 91191 Gif-sur-Yvette C\'edex, France}
\email{rchary,elbaz@ucolick.org}
\altaffiltext{1}{Department of Astronomy \& Astrophysics, University of California, Santa
Cruz, CA 95064}
\altaffiltext{2}{Physics Department, University of California, Santa Cruz, CA 95064}

\begin{abstract}

The mid-infrared local luminosity function is evolved with redshift to fit
the spectrum of the 
cosmic infrared background (CIRB) at $\lambda > 5$~\um\ and the galaxy counts from various
surveys at mid-infrared, far-infrared and submillimeter wavelengths.
A variety of evolutionary models provide satisfactory fits to
the CIRB and the number counts. The degeneracy in the range of models
cannot be broken by current
observations. 
However, the different evolutionary models yield
approximately the same comoving number density of infrared luminous
galaxies as a function of redshift.
Since the spectrum of the 
cosmic background at $\lambda > 200$~\um\ is quite sensitive to the evolution
at high redshift, i.e. $z > 1$, all models that fit the counts
require a flattening at $z\sim0.8$ to 
avoid overproducing the CIRB. 
About 80\% of the 140~\um\ CIRB
is produced at 0 $< z < 1.5$ while only about 30\% of the 850~\um\ background is produced
within the same redshift range. 
The nature of the evolution is then
translated into a measure of the dust enshrouded star formation rate density as a function
of redshift and compared with
estimates from rest-frame optical/ultraviolet surveys. The dust obscured star formation
rate density appears to peak at $z=0.8\pm0.1$, much sooner than previously thought,
with a value of 0.25$^{+0.12}_{-0.1} M_{\sun} yr^{-1} Mpc^{-3}$, in excellent agreement with a value
derived from extinction correction to NICMOS observations, and remains almost 
constant up to $z\sim2$. At least 70\% of this 
star formation takes place in infrared luminous galaxies with $L_{IR}>10^{11}~L_{\sun}$.
The long wavelength observations that constrain our evolutionary
models do not strongly trace the evolution
at $z > 2$ and a drop off in the dust enshrouded star formation rate density is consistent with
both the CIRB spectrum and the number counts. However, a comparison with the infrared
luminosity function derived from extinction corrected rest-frame optical/ultraviolet observations of
the Lyman-break galaxy (LBG) population at $z\sim3$ suggests that the almost flat comoving
star formation rate density seen between redshifts 0.8 and 2, extends up to a redshift of $z\sim4$.

\end{abstract}


\clearpage

\section{Introduction}

The Extragalactic Background Light (EBL) in the infrared, also referred to as the
Cosmic Infra-Red Background (CIRB), is a record of
the emission, absorption and re-radiation of photons integrated over cosmic history.
It provides a valuable constraint on theories of 
galaxy formation and evolution. The EBL
at near-infrared wavelengths is due to redshifted radiation from stars.
At mid-infrared (MIR) wavelengths, the background is due to
redshifted emission from dust that consists of the polycyclic aromatic
hydrocarbon (PAH) features and very small grains transiently heated to $T\sim200~K$ in
individual galaxies. At far-infrared (FIR) wavelengths, the dominant
contributor is thought to be cold dust ($T\sim20~K$) that is heated by the ambient interstellar
radiation field in galaxies.
The recent detection of this background at 2.2~\um, 3.5~\um, 140~\um\ and 240~\um\ using
DIRBE/COBE data, and in the 125$-$2000~$\mu m$ range using FIRAS/COBE measurements by
various groups \citep{Pug96, DA98, Fix98, HAKDO98, SFD98, lag99, GWC00, WR00, ELW2000}, has indicated
that the intensity of the optical/near-infrared background is
roughly equal to that of the far-infrared background. This implies
that about 50\% of the integrated rest-frame optical/ultraviolet (UV) emission from stars
and other objects is thermally
reprocessed by dust and radiated at mid- and far-infrared wavelengths. 
Thus, star formation rates that are derived from rest-frame
optical/UV luminosities of galaxies
are a lower limit to the true star formation rate
\citep[e.g.][]{Mad98, Meu99, St99, Yan99}.

The first good evidence of this came from the IRAS sky survey which revealed a new
population of galaxies with $L_{IR}=L(8-1000~\mu m) \ge 10^{11} L_{\sun}$ \citep[see review by][]{San96}.
Those with $L_{IR} \ge 10^{12} L_{\sun}$ were classified as 
ultraluminous infrared galaxies (ULIGs) while galaxies with
$10^{12} > L_{IR} \ge 10^{11} L_{\sun}$ were classified as luminous infrared galaxies (LIGs)
\footnote{Previously, the term LIG was used for all objects with $L_{IR} \ge 10^{11} L_{\sun}$. We 
use ``infrared luminous galaxies'' when referring to both LIGs and ULIGs collectively.}.
These objects exhibited the largest known star formation rates of all local galaxies
but had $\sim$90\% of the bolometric luminosity being emitted in
the far-infrared (40$-$500~\um) indicating that dust reprocessing is a significant parameter 
that needs
to be considered in estimates of star formation in certain galaxies \citep[e.g.][]{Soi86}.  
However, in 
the local universe, the integrated bolometric luminosity density
of ``normal'' optically selected
galaxies is $L_{Bol}=4\times 10^{8}~L_{\sun}~Mpc^{-3}$ while that of 
infrared luminous galaxies is $\sim8\times 10^{6}~L_{\sun}~Mpc^{-3}$
i.e. 50 times less \citep{Soi87}. This seemed to indicate that the 
contribution from LIGs and
ULIGs is sufficiently small that they need to be considered only as extreme cases.

Spectroscopic follow-up of the faint IRAS population which covered
a relatively small redshift range ($z<0.27$) indicated that 
infrared luminous galaxies were more numerous in the past than they are today and
may have a significantly larger contribution to the integrated luminosity 
density than inferred from observations of the local universe \citep{Kim98}.
Deeper observations that trace the far-infrared luminosity of
galaxies to high redshift are difficult since cirrus and confusion
noise rapidly begins to dominate.

The ISOCAM guaranteed time
extragalactic surveys in conjunction with the ELAIS survey and observations of the 
lensing cluster Abell~2390 covered a range of flux densities 
between 50~$\mu$Jy and 50~mJy at 15~$\micron$ \citep{Alt99, Elb99, Ser00}. The differential
counts resulting from these surveys revealed that the counts
of galaxies increase quite rapidly as $S_{\nu}^{-3}$ at brighter
flux levels ($S_{\nu} > 0.4$ mJy) and then flatten out as $S_{\nu}^{-1.6}$ at fainter levels.
The observed mid-infrared counts are an order of magnitude higher than expected if the local
mid-infrared luminosity function were not evolving with redshift.
This rapid increase in mid-infrared luminous
galaxies has been modeled as a 
$(1+z)^{4.5}$ luminosity evolution in the 15~\um\ local luminosity function by
\citet{Xu00}, and as a combination of number density and luminosity
evolution by \citet{Fra00}. This evolution is much stronger than
observed in the UV by \citet{Cow99} who
find that the comoving UV luminosity density evolves 
as $(1+z)^{1.5}$, instead of the $(1+z)^{3.9\pm0.75}$ 
initially proposed by \citet{Lil96}.
Furthermore, observations of galaxies in the local universe have shown that the mid-infrared
and infrared luminosities are well correlated (Section 2). The 
mid-infrared luminosity of $\sim 70$\% of the
sources seen in the ISOCAM surveys, particularly in the Hubble Deep Field-North and flanking fields
(HDFN+FF)
translates to an infrared luminosity greater than $10^{11}~L_{\sun}$ implying
that the majority of them are LIGs and ULIGs \citep{Elb00}. 
At $z\sim 0.8$, the mid-infrared luminosity density derived from the 
ISOCAM 15~\um\ sources is 
$\sim$7$\times$10$^{7}~L_{\sun}~Mpc^{-3}$ while the $8-1000~\mu m$ luminosity density adopting the
mid to far-infrared correlation seen in the local universe is 
$\sim$5$\times$10$^{8}~ L_{\sun}~Mpc^{-3}$. In comparison, at $z\sim$0, 
LIGs and ULIGs contribute $10^{6}$~$L_{\sun}~Mpc^{-3}$ to the 12~\um\ 
and 15~\um\ luminosity density and 7.8$\times$10$^{6}$~$L_{\sun}~Mpc^{-3}$
to the infrared luminosity density, as derived
from the local luminosity function (LLF) of \citet{Soi87, Fan98, Xu98}.
This indicates an increase by a factor of about 60 between $z\sim$0 and $z\sim$0.8,
providing further evidence for an evolution in the infrared luminosity function (IRLF) with redshift.

Similar deep surveys have been conducted at 850~$\mu$m\ using the SCUBA instrument on the 
JCMT \citep{Hug98, Bar99, Bl99, Eal00}. The large beam size (14$\arcsec$ FWHM) and the 
negative $k$-correction in this wavelength regime makes identification of the optical counterparts
and thereby the redshift distribution of the sources very difficult. High resolution radio 
interferometric observations and the use of 450~\um/850~\um\ flux ratios have helped somewhat
in localizing the sources and constraining the redshifts \citep{Hug98, bar00}. These have placed
the bright ($S_{\nu} > 6$~mJy)
submillimeter sources at $z\sim1-3$, which has been confirmed by the more extensive survey
of \citet{Chap00}. The implication of this is that most, if not all the submillimeter sources
are extreme ULIGs with star formation rates of $10^{2}-10^{3}~M_{\sun}~yr^{-1}$. Furthermore,
the star formation rate density due to ULIGs must have increased by about
2 orders of magnitude between $z\sim$0 and $z\sim 1-3$. 

Many of the LIGs and ULIGs in the local
universe show morphological signatures of interaction, and
$>$50\% of the optical counterparts of ISOCAM HDFN
galaxies show evidence for interactions \citep{Mann97}.
Surveys at visible wavelengths show a redshift evolution of the merger fraction,
defined as the fraction of close pairs of galaxies,
as $\sim (1+z)^{3}$ \citep[e.g.][]{Lef00}. Thus, if mergers were indeed a tracer of LIGs
and ULIGs, this would again suggest that the bright end of the IRLF is evolving strongly.
However, the faint end of the IRLF is very
poorly constrained at $z\sim1$ since none 
of the long wavelength surveys are sensitive enough to detect
galaxies with $L_{IR} < 10^{11}~L_{\sun}$ at $z > 0.5$. \citet{Meu99} has shown that
the FIR to UV flux ratio is closely related to the
UV slope for normal starbursts but that the relationship breaks down for ULIGs \citep{Meu00}.
This indicates that the visible/near-infrared counts can potentially place constraints
on the evolution of the faint end of the infrared luminosity function but we postpone
this discussion to the future.

In this paper, we combine data from a variety of published surveys of nearby galaxies
to determine the correlation, if any, between the luminosities at various mid- and far-infrared
wavelengths. We use these correlations to generate smoothly varying spectral
energy distributions for galaxies as a function of luminosity class. We 
assess the need for luminosity and density evolution in the 15~\um\ luminosity function
of \citet{Xu98} and therefore the 60~\um\ luminosity function of 
\citet{Soi87} based on 
fits to the ISOCAM 15~\um, ISOPHOT 90~\um, ISOPHOT 170~\um\ and SCUBA 850~\um\ galaxy counts,
as well as the spectrum of the CIRB at $\lambda > 5$~\um. 
The evolution of the mid-infrared luminosity function
is then translated to an estimate of the dust enshrouded star formation rate density (SFR) 
as a function of redshift
and compared with SFR values derived from optical/near-infrared surveys.
We adopt a $H_{0}$=75~km~s$^{-1}$~Mpc$^{-1}$,
$\Omega_{M}=0.3$, $\Omega_{\Lambda}=0.7$ cosmology throughout this paper unless otherwise
explicitly stated. 

\section{Luminosity Correlations in the Infrared and Template Spectral Energy Distributions}

\ifpp
\begin{figure}[ptb]
\plotone{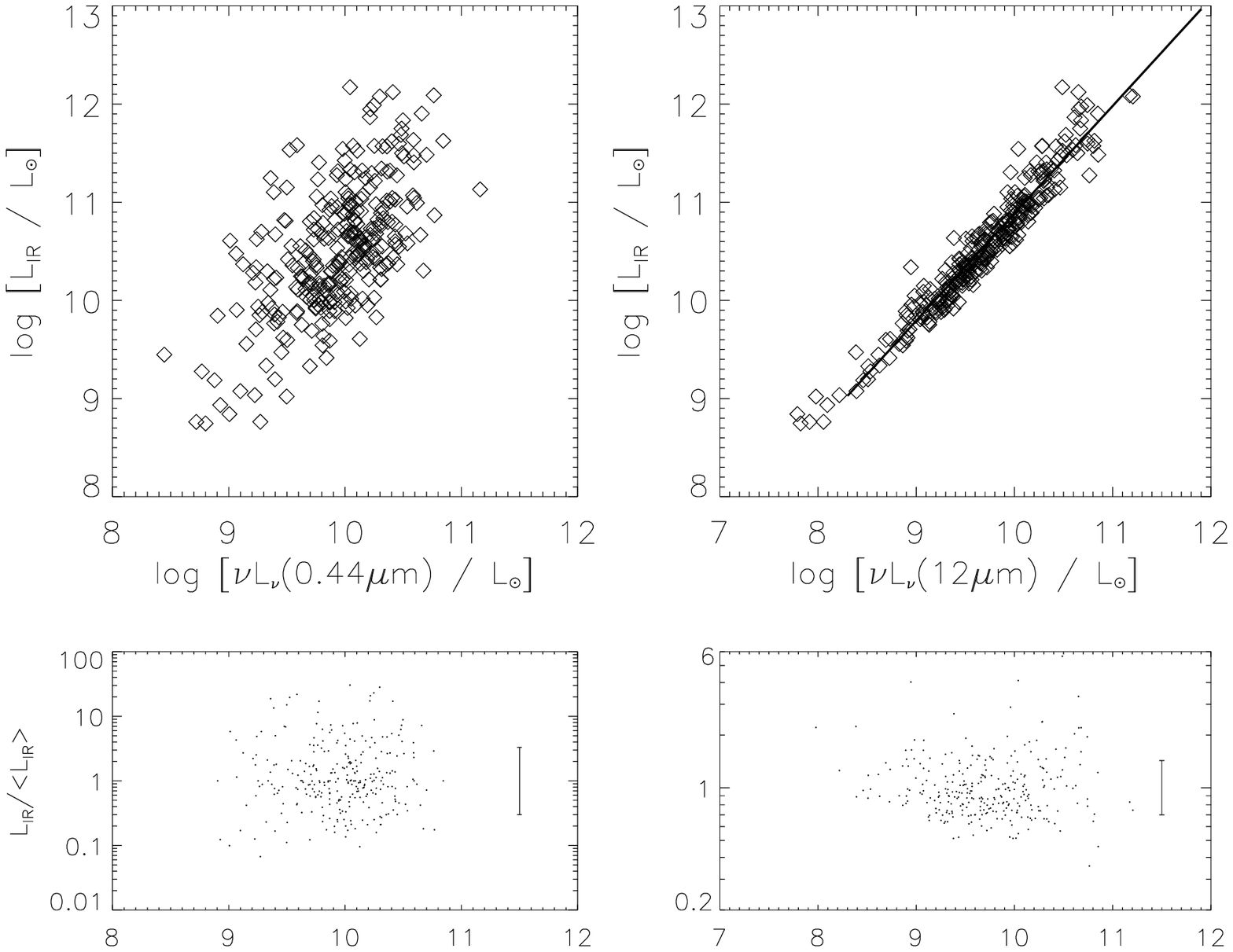}
\caption{
Plots showing the relative accuracy of tracing 
infrared luminosities (8$-$1000~\um) of IRAS BGS \citep{Soi87}
objects from the $B-$band (0.44~\um) and 12~\um\ luminosities.
The lower plots show the ratio between true infrared luminosity ($L_{IR}$) and 
predicted infrared luminosity ($<L_{IR}>$) derived from the $B-$band or 12~\um\ luminosity using
a first order polynomial fit to data points
with $L_{IR} > 10^{10}~L_{\sun}$. If the infrared luminosity of
all galaxies could be predicted precisely from their $B-$band or 12~\um\ luminosity, then
all points in the lower plot would lie in a horizontal line with $L_{IR}/<L_{IR}>$=1.
The lower plots also show the 1$\sigma$ uncertainty
in the prediction of the infrared luminosities.
}
\label{fig:mf1}
\end{figure}
\fi

\ifpp
\begin{figure}[ptb]
\plotone{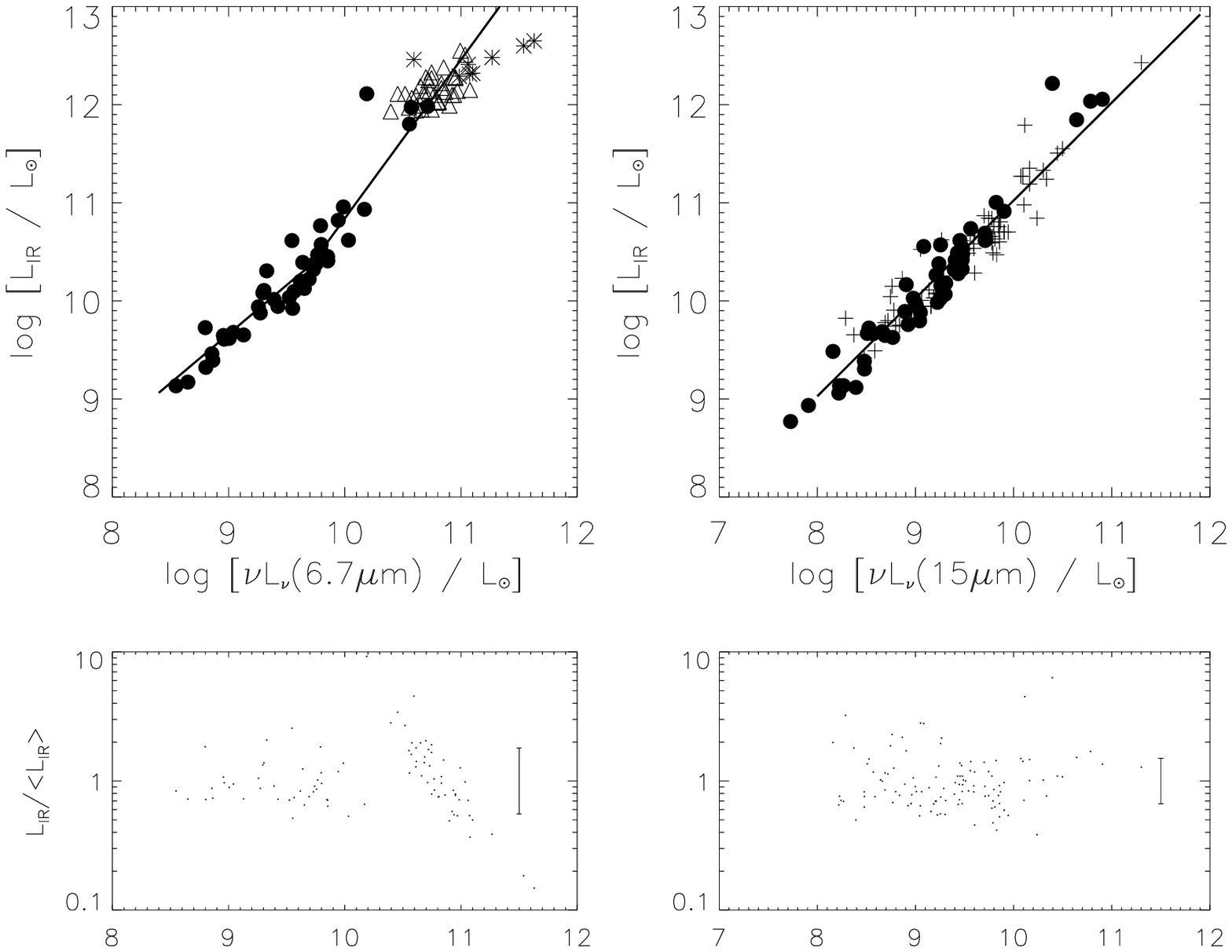}
\caption{
Plots showing the relative accuracy of tracing
infrared luminosities (8$-$1000~\um) from the 6.7~$\mu$m and 15~\um\ luminosities.
The 6.7~$\mu$m luminosities are from ISOCAM guaranteed time surveys \citep[filled circles;][]{Chan00}
and ISOPHOT 7~$\mu m$ spectroscopy of ULIGs (triangles) by \citet{Rig99}.
Note that the
\citet{Rig99} values have been modified as described in the text.
Asterisks
are the 6.7~\um\ luminosities for the starburst dominated ULIG sample of \citet{Tra01}.
Some of the extreme ULIGs might have a significant AGN contribution which could result in a
deviation from the 6.7~\um-to-IR correlation derived for starbursts by decreasing the IR luminosity for
a given 6.7 luminosity, as for the two brightest galaxies
in our sample.
The 15~\um\ luminosities are from the \citet{Chan00} (filled circles) and the \citet{Aus00} (pluses) sample of galaxies.
The lower plots are similar to those in Figure \ref{fig:mf1}.
}
\label{fig:mirfir}
\end{figure}
\fi

It can be shown that the 12~\um\ and far-infrared luminosities
of galaxies in the IRAS Bright Galaxy Sample (BGS) cannot be accurately 
derived from their $B-$band
luminosities \citep{Soi87}\footnote{The Zwicky magnitudes,
$m_{z}$, in the BGS were converted to B-luminosities using $m_{B} = m_{z}-0.14$ and 
a $B-$band zero point of 4260~Jy.}. The peak to peak scatter in the $L_{B}/L_{IR}$
ratio for a fixed $L_{IR}$ is about a factor of 20.
However, as mentioned earlier, the FIR to UV flux ratio has been 
shown to be closely related to the
UV slope for normal starbursts \citep{Meu99}. This relationship breaks down for the
ULIGs \citep{Meu00}. The phenomenon can be qualitatively explained by the fact that the UV emission
arises from stars which are relatively unobscured to the observer. Regions of star formation
with large optical depth i.e. \ion{H}{2} regions, could exist 
in the central regions of galaxies where almost all the
UV light is reprocessed to the MIR and FIR. 
Thus, the regions of FIR and UV emission would be unrelated, especially for ULIGs.
The best observational evidence for this explanation can 
be seen in the Antennae galaxy \citep{Mir98}
where about half the 15~\um\ emission seen by ISOCAM arises from regions that are
inconspicuous at visible wavelengths. The breakdown in the FIR to UV slope correlation for ULIGs
is problematic for determination of the true star formation rate from optical/UV surveys, 
since submillimeter observations using the SCUBA instrument
indicate that ULIGs might have a larger contribution
to the star formation rate density at high redshift. This suggests that
 it will be difficult to determine the true
SFR by applying an accurate extinction correction to the optical/UV determined value.

Since the short wavelength starlight and dust emission are not closely related,
an estimate of the dust enshrouded star formation rate can only be derived from
other tracers such as 
the mid-infrared and far-infrared luminosities or using the radio to far-infrared correlation
shown by \citet{Con92}. 
In the mid-infrared
regime, the spectra of galaxies exhibit broad emission features at 6.2, 7.7, 8.6, 11.3
and 12.7~\um\ which are probably from
polycyclic aromatic hydrocarbons \citep[see review by][]{pug89}. 
These features and their associated continuum dominate the emission at mid-infrared wavelengths shortward
of 10~\um. There is, 
in addition, a continuum 
from very small grains (VSGs) of size $<$10~nm which dominates the emission
above $\sim$10~\um\ \citep{des90, lau00} except for quiescent star forming galaxies.
The VSGs get transiently heated to
temperatures of $\sim$200~K by the ambient optical/UV continuum which is proportional to the
star formation activity. 
In addition, mid-infrared measurements do not need large
extinction corrections since the extinction at mid-infrared wavelengths is only
about 1\% that at visible wavelengths \citep{mat90}. The radio wavelengths on the other hand
are dominated by free-free
emission from HII regions and synchrotron emission from supernova remnants.
Although radio observations are almost confusion limited at an 8.5 GHz 
sensitivity of 9~$\mu$Jy obtained over the HDF \citep{Rich98}, 
we find that they are typically as sensitive as the ISOCAM 15~\um\ observations in that
they can typically probe galaxies with $L_{IR}\sim10^{11.4}~L_{\sun}$ at $z\sim1$. 
Since the ISOCAM 15~\um\ observations provide the primary constraint on evolution
models at $z < 1.2$, we adopt as a starting point the 
local 15~\um\ luminosity function described in \citet{Xu98} and \citet{Xu00}. For the rest of this paper, we will use
the convention defined in \citet{San96}:
\begin{equation}
L_{IR} = L(8-1000~\mu m) = 1.8\times10^{-14}\times10^{26}~[13.48 L_{12} + 
5.16 L_{25} + 2.58 L_{60} + L_{100}] 
\end{equation}
\begin{equation}
L_{FIR} = L(40-500~\mu m) = 1.6\times1.26\times10^{-14}\times10^{26}~[2.58 L_{60} + 
L_{100}]
\end{equation}
\noindent In the above, the symbol $L_{\lambda}$ is defined as $L_{\nu}(\lambda~\mu m)$ in units
of $L_{\sun}~Hz^{-1}$. $L_{IR}$ and $L_{FIR}$ are in $L_{\sun}$.

To use the mid-infrared luminosity function as a tracer of the dust enshrouded star formation
rate, we first need to define a calibration scale. Figures \ref{fig:mf1} and \ref{fig:mirfir} 
illustrate
the accuracy with which the infrared luminosity of galaxies can be derived from their
mid-infrared luminosities. Figure \ref{fig:mf1} is based on $\sim$300 galaxies from the
IRAS BGS while Figure \ref{fig:mirfir}
is based on published ISOCAM and ISOPHOT observations of IRAS galaxies.
The data points with $L_{IR}>10^{10}~L_{\sun}$ are fit by a first order
polynomial shown in Equations 4-6. This is shown as a solid black line in the upper panels
of the figure. Objects with
$L_{IR} < 10^{10}~L_{\sun}$ are not used for the fits. At these low luminosities, the fraction of
the bolometric luminosity emitted in the infrared is $<$50\% i.e.
visible starlight that is not obscured by dust is dominating the radiated energy of
the galaxy. On the other hand a significant fraction of the more luminous objects show
disturbed morphologies, suggesting interactions with other galaxies that would result
in gas-rich systems with star formation in highly obscured regions.
The lower plots in both figures show the scatter in the ratio of the 
infrared luminosity as derived from IRAS data for these galaxies to the infrared
luminosity derived from the polynomial fits.
Also shown is the 1$\sigma$ uncertainty in the derived infrared luminosity calculated as
the range within which 68\% of the galaxies lie. The lowest luminosity 
objects ($L_{IR}<10^{9}~L_{\sun}$) have been rejected in the lower plots.

The 6.7~\um\ luminosities
were derived from ISO observations of $\sim$90 nearby starburst dominated
galaxies. 44 spiral and starburst galaxies had photometry from ISOCAM
\citep{Chan00, Rouss01, lau00},  8 ULIGs had spectra from 
ISOCAM circular variable filters (CVF) observations \citep{Tra01} while
37 ULIGs had ISOPHOT mid-infrared spectra \citep{Rig99}. 
\citet{Rig99} obtained mid-infrared spectra with ISOPHOT of about 60
ULIGs and about 15 low luminosity starbursts and normal galaxies to study the emission features
from the PAHs. Of the 60 ULIGs, about 45 had the 7.7~\um\ PAH
feature detected with good signal to noise. However,
the calibration and performance of the instrument is not very well determined.
To assess the quality of the data set, we compared the ISOPHOT observations
on 5 ULIGs and 4 low luminosity starbursts/normal galaxies to the \citet{Chan00} ISOCAM LW2 observations of 
the same galaxies\footnote{The LW2 filter is broad enough to include the 7.7~\um\ PAH feature but
is centered at 6.75~\um.}. 
We find that for the ULIGs, the ratio of 7.7~\um\ line+continuum
flux density as published in Rigopoulou et al. to the ISOCAM  6.7~\um\ flux density lies
in the range 1.5$-$3.0. In comparison, for the starbursts/normal galaxies,
the ratio of 7.7~\um\ line+continuum to the ISOCAM  6.7~\um\ flux density falls
in the range 0.2$-$1.6, a factor of 8. 
So, for assessing the correlation
between the mid and far-infrared luminosities, we consider the ISOCAM data
on bright IRAS galaxies
as well as the ULIG sample of \citet{Rig99}, dividing
the line+continuum flux value published in the latter by 2.4 and assigning
a peak to peak error bar of a factor of 2. This is consistent with the range of
1.5$-$2.7 that we find for the 7.7~\um\ line+continuum to the 6.7~\um\ flux density
ratio in the ISOCAM CVF observations of \citet{Tra01}.

The 15~\um\ and infrared luminosities of 120 IRAS galaxies were taken from
the sample of \citet{Chan00}
and the survey performed in the North Ecliptic Pole Region (NEPR) by \citet{Aus00}. 
The NEPR sample of galaxies only has IRAS 60~\um\
luminosities available and we have 
converted these to a far-infrared luminosity
based on a 60~\um\ to far-infrared correlation derived by combining the IRAS BGS and the
IRAS PSCz catalog of \citet{Sau00}. 
The far-infrared luminosity is typically about 83\% of the
total infrared luminosity and we have applied this conversion to be consistent with the
other plots.

Clearly, the 6.7~\um, 12~\um\ and 15~\um\ luminosities of galaxies
trace the infrared luminosity much better than the $B-$band luminosity.
The 15~\um-to-IR and 12~\um-to-IR correlations were determined
from a first order polynomial fit to the data with $L_{IR}>10^{10}~L_{\sun}$.
Since the 6.7~\um-to-IR correlation is based on different data sets from
CVF and broadband photometry with ISOCAM and spectroscopy using ISOPHOT,
the polynomial fit for the 6.7~\um-to-IR correlation
was determined by applying a $k-$correction from 6.7~\um\ to 15~\um\
based on ISOCAM observations of nearby galaxies (Figure \ref{fig:compcorr})
and then using the 15~\um-to-IR correlation. 
The mid-infrared to IR correlations 
show a similar scatter around the correlation line which is about a factor of 5
better than the optical-to-IR correlation. 

These data sets tentatively illustrate the potential of using the 
mid-infrared as a tracer of dust
enshrouded star formation and a more homogeneous and comprehensive survey of nearby galaxies,
as will be undertaken by SIRTF, will be required to either strengthen or reject this correlation.

\citet{Kenn98} has transformed the infrared 
luminosity of young $<10^{8}$ yr starburst galaxies to 
a star formation rate.
If we adopt the correlations shown in the previous figures, we 
can translate the mid-infrared luminosity of galaxies with $L_{IR}>10^{10}~L_{\sun}$
to an approximate estimate of the dust enshrouded SFR ($\rho '$) using the
formulae:

\begin{eqnarray}
\rho ' (M_{\sun}~yr^{-1}) & = & 1.71\times10^{-10}~L_{IR}~(L_{\sun}) \\
L_{IR} & = & 11.1^{+5.5}_{-3.7} \times L_{15~\mu m}^{0.998} \\
L_{IR} & = & 0.89_{-0.27}^{+0.38} \times L_{12~\mu m}^{1.094} \\
L_{IR} & = & 4.37^{+2.35}_{-2.13} \times10^{-6} \times L_{6.7~\mu m}^{1.62} 
\end{eqnarray}

\noindent where all values are in $L_{\sun}$.
The 1$\sigma$ values 
have been estimated
by calculating the range of values within which 68\% of galaxies have their 
observed infrared luminosities. 

As mentioned earlier, the main observational constraints on models that trace the
redshift evolution of the IRLF are:
\begin{enumerate}
\item differential counts from various surveys at mid-infrared, far-infrared
and submillimeter wavelengths and
\item the spectrum of the CIRB at $\lambda > 5$~\um.
\end{enumerate}
To use these constraints, it is necessary to know the
luminosity at different wavelengths for galaxies in each luminosity bin of the infrared 
luminosity function.
This motivates the generation of template spectra for objects of different luminosity classes.
It is useful to note that many evolutionary models which have already been
developed either use a mid-infrared template which is a very poor representation
of the true PAH emission features \citep[e.g.][]{Row00, Pea01, Sad01} or neglect the PAH features
altogether \citep{Mal98, Mal00}. It is trivial to show that this makes a
critical difference in the quality of the fits to
the ISOCAM mid-infrared number counts and therefore affects the evolution parameters
particularly at $z<1$. So, it is important to have template SEDs that reproduce the observed trend
in the luminosity of local galaxies at different wavelengths.

Using MIR, FIR and submillimeter
data from ISOCAM, IRAS and SCUBA observations of nearby galaxies,
we fit the observed trend between different mid- and far-infrared luminosities as
shown in Figure \ref{fig:compcorr}. The 
top two panels in the figure show the
ISOCAM observations at 6.7~\um\ and 15~\um\
for $\sim$50 IRAS galaxies which are described in \citet{Chan00}. 
The solid lines in the panels show two first order polynomial fits, one for galaxies 
with 15~\um\ luminosity
less than $2\times10^{9}~L_{\sun}$ and another for more luminous galaxies. This is because
the ratio
between the mid-infrared luminosities changes as a function of 15~\um\ luminosity, possibly
due to enhanced emission from the VSG component \citep{lau00}. The luminosity break corresponds to $L_{IR}\sim
2\times10^{10}~L_{\sun}$ which is similar to the luminosity cutoff used for deriving Equations 4-6.
The panel showing the 15~\um\ to 60~\um\ trend consists of data described in 
Figure \ref{fig:mirfir}. The 60~\um\ to FIR correlation is for the IRAS BGS and PSCz
galaxies while the panel showing the IR to FIR correlation is only for the IRAS BGS galaxies. 
The last panel shows SCUBA submillimeter data 
on $\sim$100 IRAS galaxies \citep{dun00}. 
In addition,
the correlation between IRAS 25~\um\ and 100~\um\ luminosities for galaxies in the BGS 
was also determined.
The solid lines for the four lower panels
utilize only a single first order polynomial fit to all the data points. 
Also shown in
the panels as red triangles are the luminosities at 
the corresponding wavelength for the different templates
that were generated as described below. 

\ifpp
\begin{figure}[ptb]
\plotone{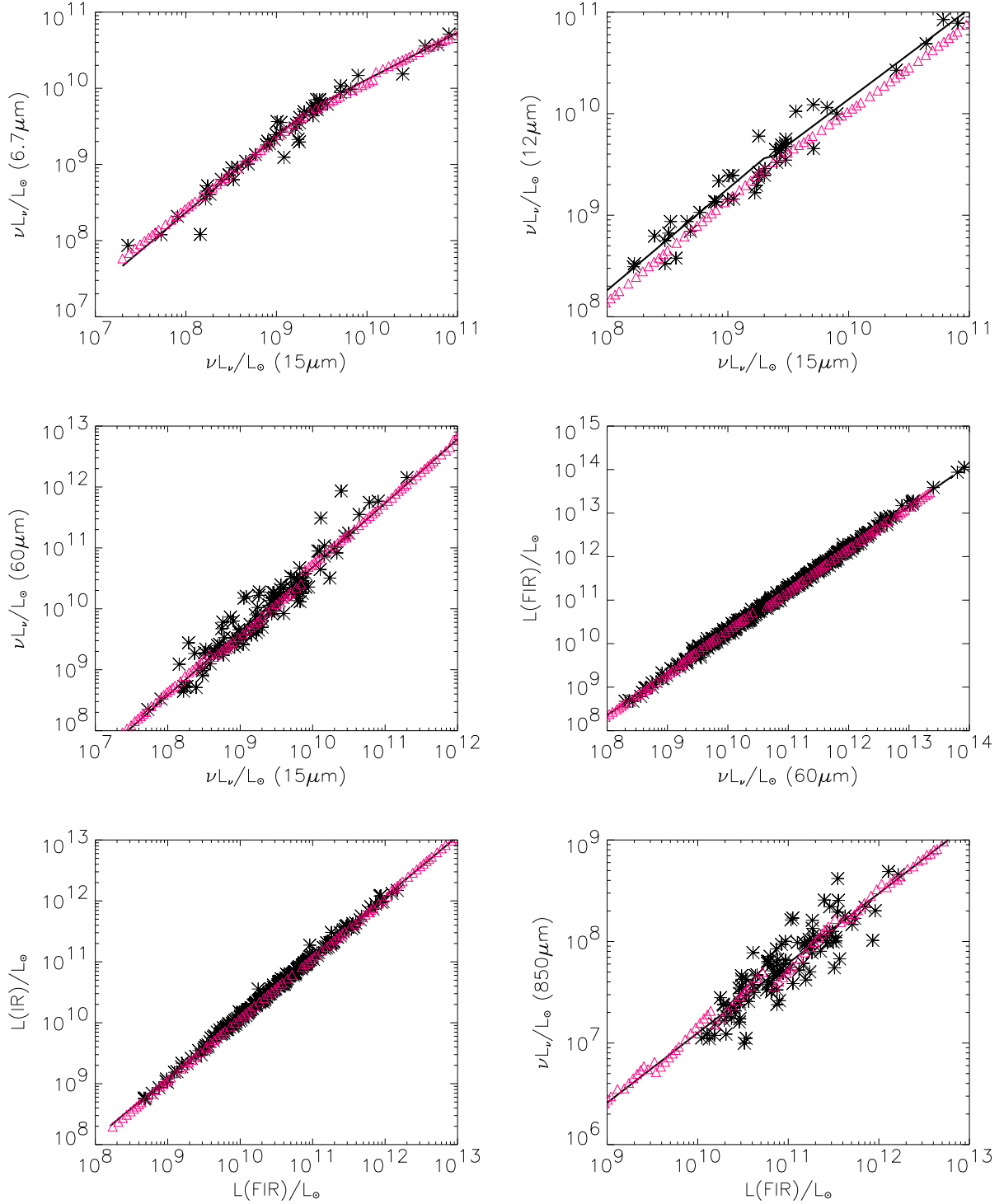}
\caption{Plot showing the data (asterisks) at different wavelengths from IRAS, ISOCAM
and SCUBA surveys \citep{Soi87, Aus00, dun00, Sau00, Chan00}. 
The lines are the best fit polynomial of order 1. The red 
triangles are the corresponding values from our template spectral energy distributions
which were generated as described in the text.
}
\label{fig:compcorr}
\end{figure}
\fi

Template spectral energy distributions (SEDs) were generated 
between 0.1 and 1000~\um\ to reproduce the observed trend
between mid-infrared and far-infrared luminosities. 
To generate these templates,
we used the basic \citet{Sil98}
models to reproduce the ultraviolet-submillimeter SED
of 4 prototypical galaxies: Arp220, NGC6090, M82
and M51. These correspond to 
objects of four different luminosity classes - ULIGs, LIGs, ``starbursts'' (SB) 
and ``normal galaxies''
respectively.
ISOCAM CVF observations between 3 and 18~\um\ of these galaxies provided new data on the relative
strength of the mid-infrared features and continuum \citep{Char99, lau00, for01, Rouss01}. 
The mid-infrared region of the modelled spectra
were then replaced with the ISOCAM observations. In 
addition, corrections were made for the 17.9~\um\ silicate
feature based on observations by \citet{Smi89}.
The four template spectra were checked to ensure that the IRAS
observed values of these four galaxies were reproduced. We then
partitioned the four templates into a mid-infrared (4$-$20~\um) and far-infrared (20$-$1000~\um)
component and interpolated between the four to generate a range of mid-infrared
and far-infrared sample templates of intermediate luminosity.
An additional set of far-infrared templates provided by \citet{Dale00} were added to the
ensemble of far-infrared templates to span a wider range of spectral shapes. 

For each luminosity bin of the 15~\um\ luminosity
function, the luminosity at the following wavelengths: 6.7, 12, 25, 60, 100, 850~\um\
was predicted based on the polynomial fits to the data shown in Figure \ref{fig:compcorr}.
Of the $\sim100$ mid-infrared sample templates generated as described above, 
the mid-infrared template that best fits the predicted 6.7, 12 and 15~\um\ luminosities was selected.
Similarly, the far-infrared template that best fits 
the predicted 25, 60, 100 and 850~\um\ luminosities was selected.
The luminosity of the templates at the corresponding
wavelengths was determined by integrating over the filter
curves of the instruments. 
Our goal was only to generate SEDs that
reproduce the observed trend in luminosities at different wavelengths. Selecting
a variety of sample templates provided better fits to the predicted luminosities
than by just interpolating
between the four SEDs generated by the \citet{Sil98} models.
The best fitting mid-infrared and far-infrared templates were
then merged together to provide the final template SED for each luminosity bin.
The red triangles in Figure \ref{fig:compcorr} are the luminosities at the corresponding
wavelengths from the final merged template SEDs. 
The $B-$band luminosity of galaxies
in the IRAS BGS
shown in Figure \ref{fig:mf1} was also used to constrain the
optical/near-infrared spectral energy distribution of galaxies but as stated before, we have
not constrained the UV slope of the template SEDs. The absence
of a good correlation between the $B-$band and IR luminosities implies that the 
optical/near-infrared part of our SEDs is highly uncertain. This is not a major problem
since we are only analyzing the dust emission in this paper.
The templates for three objects with infrared
luminosities of $10^{10}, 10^{11}$ and $10^{12}~L_{\sun}$ along with the predicted luminosities
at different wavelengths based on the correlations in Figure \ref{fig:compcorr}
are shown in Figure \ref{fig:template}.

\ifpp
\begin{figure}[ptb]
\centerline{\psfig{file=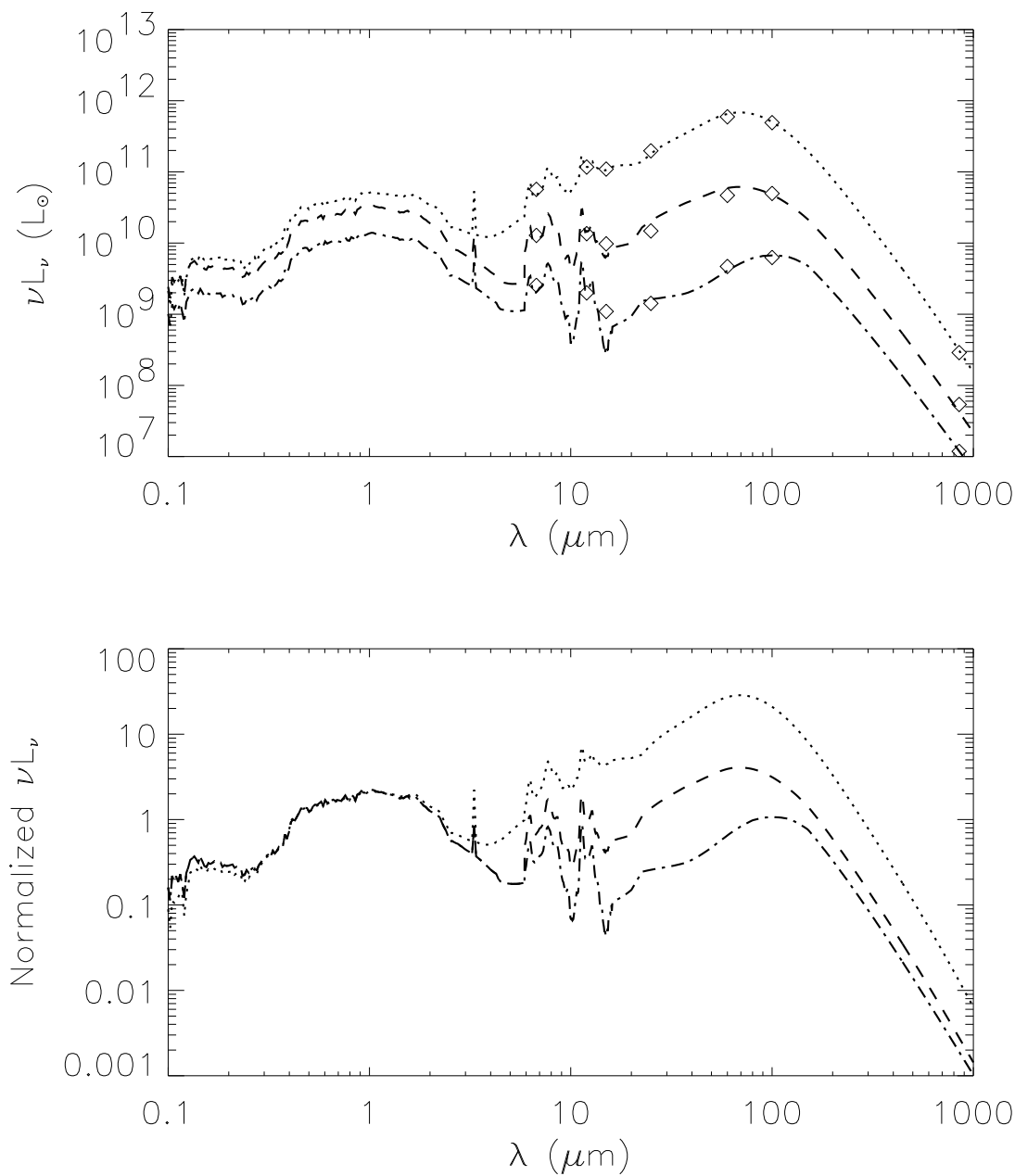,height=7.0in}}
\caption{
Template spectral energy distribution for objects of three different infrared luminosities
along with the predicted luminosities at different wavelengths (diamonds). The luminosities
correspond to $L_{IR}=10^{12}, 10^{11}$ and $10^{10}~L_{\sun}$ illustrating the SED of
ULIGs, LIGs and starbursts respectively. The lower plot
shows the same templates, normalized at 0.44~\um\ ($B-$band) to show the evolution of the spectrum
as a function of infrared luminosity. No correction for the UV slope has been made.
}
\label{fig:template}
\end{figure}
\fi

\section{Evolution of the 15~\um\ and Far-Infrared Local Luminosity Function}

Since our intention is to use the different mid- and far-infrared
observational constraints to estimate the evolution of the dust enshrouded star 
formation rate with redshift, we use the 15~\um\ LLF as a tracer 
of dust emission in the local universe.
\citet{Xu98} and \citet{Xu00} derived a 15~\um\ LLF
based on a correlation between ISOCAM mid-infrared and IRAS mid- and far-infrared data.
In addition, estimates of the 12~\um\ LLF have been made by \citet{Rush93} and \citet{Fan98}.
The different 12~\um\ and 15~\um\ luminosity functions 
are shown in Figure \ref{fig:llf}. Also shown is the predicted 60~\um\ LLF derived from the 
mid-infrared LLF using a mid-infrared to 60~\um\ conversion from the 
polynomial fit to the observations
of \citet{Aus00} and \citet{Chan00} described earlier.
All these are in good agreement with each other since they were essentially
derived from IRAS observations of nearby galaxies.

\ifpp
\begin{figure}[ptb]
\centerline{\psfig{file=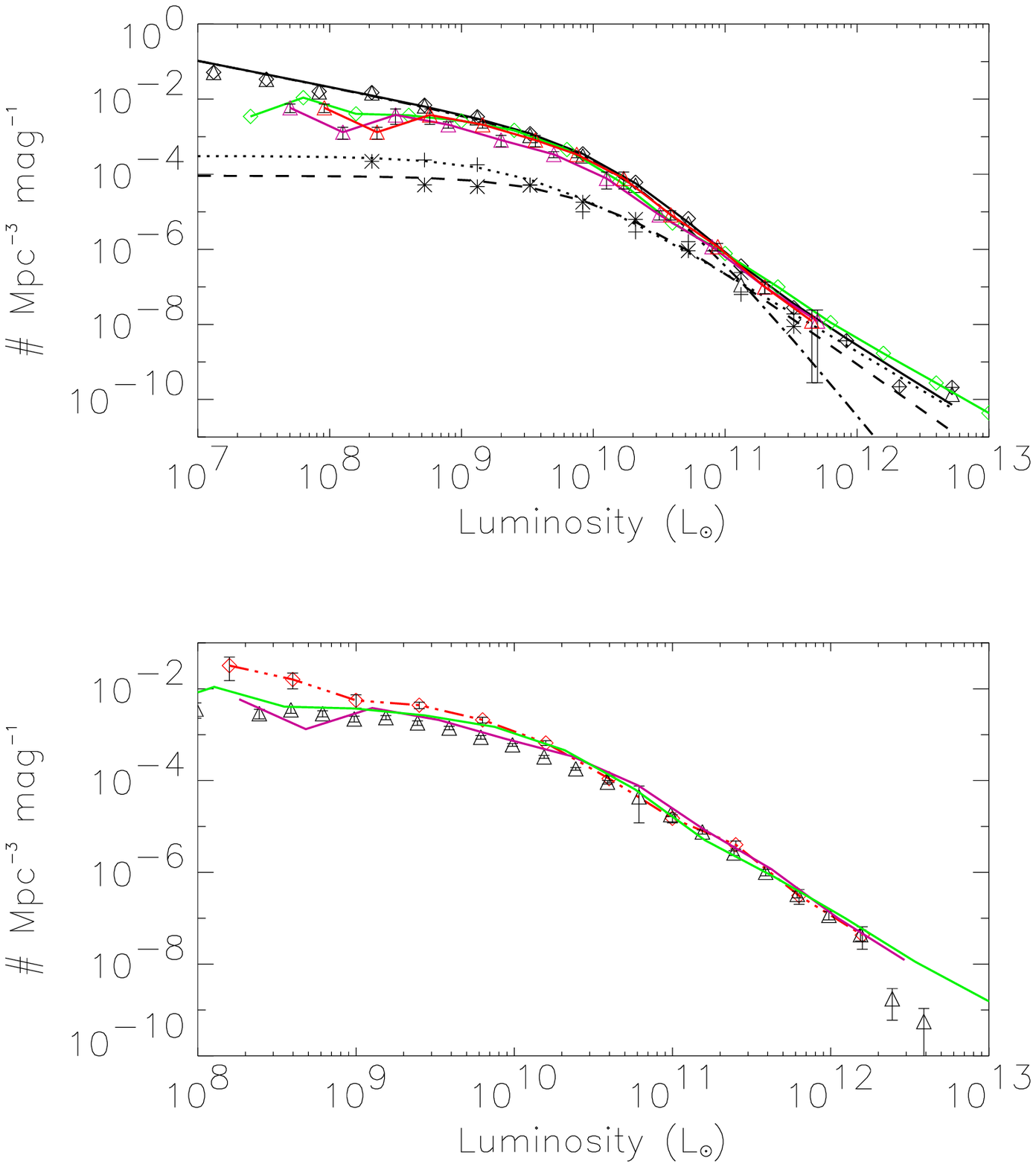,height=6.5in}}
\caption{Various 12~\um\ and 60~\um\ local luminosity functions for $H_{0}$=75 km~s$^{-1}$~Mpc$^{-1}$.
In the upper plot, black lines and symbols represent the 12~\um\ LLF from \citet{Rush93}.
The solid black line is the total LLF, black dash-dot line is the
non-seyfert contribution, dotted line is from Seyfert 1s, dashed line is from Seyfert 2s.
The green line and symbols are the 12~\um\ LLF from \citet{Fan98}.
The purple line and triangles are the 15~\um\ LLF from \citet{Xu98} and \citet{Xu00}.
The red line is this 15~\um\ LLF converted to 12~\um\ based on a $k$-correction derived
from ISOCAM observations of 44 nearby galaxies.
The lower plot shows the 60~\um\ LLF from \citet{Soi87} as the broken red line, the
60~\um\ LLF of \citet{Sau90} as the black triangles, the 12~\um\ LLF of \citet{Fan98}
converted to 60~\um\ using a linear 12~\um\ to 60~\um\ correlation based on IRAS BGS data
(green line) and the \citet{Xu98} 15~\um\ LLF converted to 60~\um (purple line).
}
\label{fig:llf}
\end{figure}
\fi

Evolution of the luminosity function with respect to redshift can be expressed as:
\begin{eqnarray}
\Psi(L,z)  & = & n(z)~\phi \left [ L,z \right ] \\
\phi(L,z)  & = & \phi \left [ \case{L}{g(z)},0 \right ]
\end{eqnarray}
\noindent where $\Psi(L,z)$ is the number density of galaxies as a function of luminosity $L$
and redshift $z$. The $n(z)$ term represents
evolution in the number density of galaxies while the $\phi [L,z]$ term represents luminosity
evolution.  $\phi [L,0]$ is the local luminosity function.
We consider models where $n(z)$ is of the form $n(0)~(1+z)^{\alpha_{D}}$ up to a turnover
redshift $z_{turn}^{D}$
followed by $n(z_{turn}^{D}) [(1+z)/(1+z_{turn}^{D})]^{\beta_{D}}$ up to $z=4.5$. The luminosity evolution
component $g(z) = (1+z)^{\alpha_{L}}$ up to $z_{turn}^{L}$
followed by $n(z_{turn}^{L}) [(1+z)/(1+z_{turn}^{L})]^{\beta_{L}}$.

It should be emphasized that there is considerable degeneracy in the density and 
luminosity evolution
of galaxies. While 
density evolution slides the luminosity function along the vertical axis,
the latter slides it on the horizontal axis. However, current observations at mid-
and far-infrared wavelengths detect galaxies only at the 
luminous end of the luminosity function, as a result of which
the two are indistinguishable. This is shown in
Figure \ref{fig:deglf}. The figure also shows that
evolving just the luminous end of the LLF
i.e. $L > 5\times10^{10}~L_{\sun}$ similar to the model proposed by \citet{Dole00}
would result in the same degeneracy with observations although it would result in
a somewhat unphysical break in the luminosity function.

There is an additional degeneracy induced by the fraction
of the luminosity function that is evolving. 
Redshift measurements of ISOCAM 15~\um\ sources in the HDFN+FF indicate that
the majority of them are LIGs and ULIGs \citep{Elb00}.
Interestingly, when the local 60~\um\ luminosity function is compared to the
Schechter function commonly used to represent the local luminosity function at visible
wavelengths, an excess of galaxies is seen in the 60~\um\ LLF beyond $L_{60~\mu m}>10^{11}~L_{\sun}$
since the Schechter function drops faster at the bright end.
On one hand it seems likely that just this excess of galaxies, most of which show morphological
signatures of merger activity, could be evolving at high redshift. Alternatively,
it is possible that a luminosity dependent fraction which
approaches 100\% at $L_{60~\mu m}>10^{11}~L_{\sun}$ of the LLF could be evolving. Unfortunately,
the observational constraints on the faint end of the IRLF are limited since
these galaxies are undetected at mid- and far-infrared wavelengths at $z>0.5$. The correlation
between mid-infrared and
visible wavelengths being poor, the counts of galaxies at visible/near-infrared wavelengths
cannot be used to constrain the distribution. However, we will investigate in a future paper
if the relationship between the FIR/UV flux ratio and UV slope can constrain the
evolution of the faint end of the IRLF. 

\ifpp
\begin{figure}[ptb]
\plotone{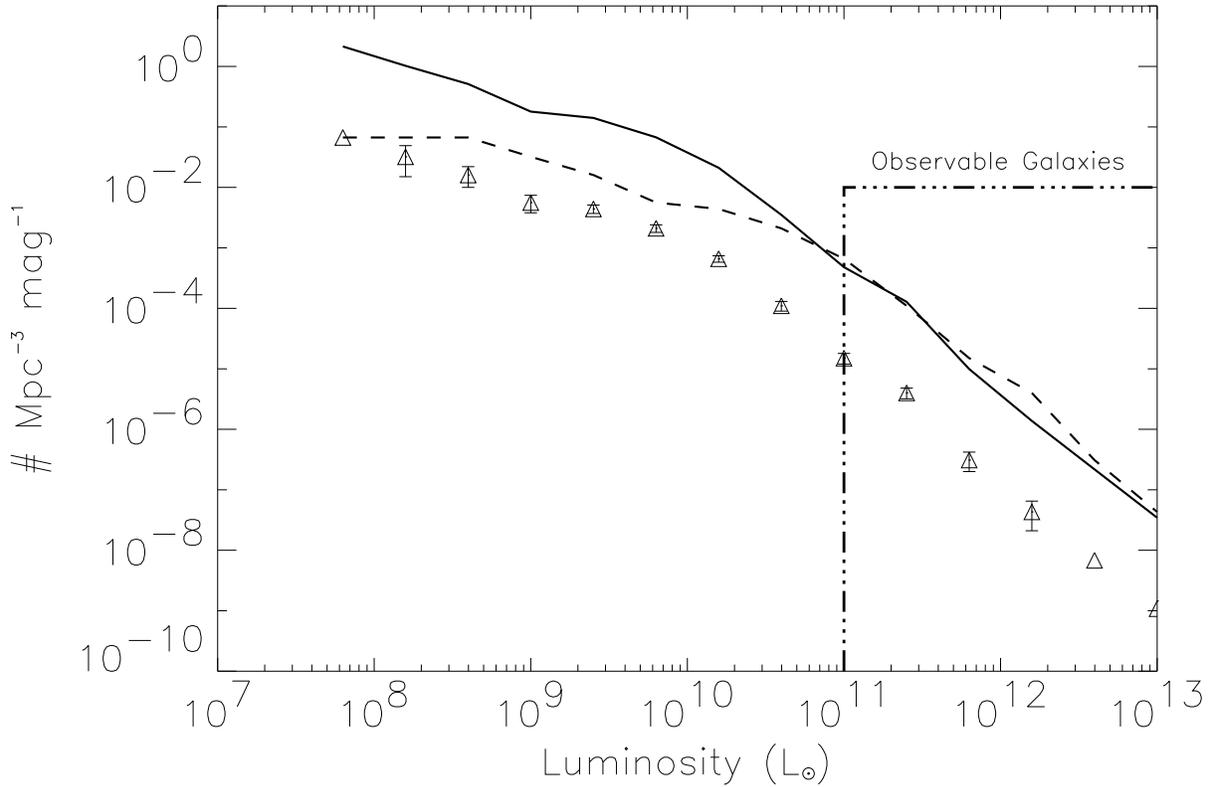}
\caption{The figure shows the degeneracy between luminosity and density evolution. The
triangles are the extrapolated 60~\um\ LLF of \citet{Soi87}, the solid line is the
resultant luminosity function at $z\sim1$ assuming a density evolution of $(1+z)^5$
while the broken line is the luminosity function at $z\sim1$ assuming a luminosity evolution
of $(1+z)^2$. Also shown is the typical luminosity range of galaxies that have been detected by
current long-wavelength surveys. 
}
\label{fig:deglf}
\end{figure}
\fi

The principal observational constraints on the evolution of the bright end of the
luminosity function then are:
\begin{itemize}
\item The ISOCAM differential number counts at 15~\um\ especially the ``knee'' in the counts 
slope at 0.4 mJy \citep{Elb99},
\item The 15~\um\ extragalactic background light which has a lower limit from
ISO counts and an upper limit from gamma-ray observations of a TeV flare
in Markarian 501,
\item The redshift distribution of 15~\um\ sources in the HDFN+FF
which is somewhat
peaked at $z\sim0.8$ \citep{Auss01} and indicates that 45\%
of galaxies with $0.1 < S_{15} < 0.4$~mJy are LIGs and 20\% are ULIGs with the remaining being
normal and low luminosity starburst galaxies \citep{Elb00},
\item The spectrum of the cosmic far-infrared background between 100 and 850~\um\
as measured by DIRBE and FIRAS
on COBE,
\item ISOPHOT 90~\um\ and 170~\um\ counts and SCUBA 850~\um\ number counts \citep{Efs00, 
Hug98, Bl99, Bar99, Eal00, Dole01}. 
\end{itemize}

The $k-$correction of galaxies illustrated in Figure \ref{kcorr}
clearly illustrates the range of redshifts that can be studied by observations at these 
wavelengths. The 850~\um\ observations are typically confusion limited at 2~mJy.
However, deeper lensed surveys or using high resolution radio interferometric data 
can push the detection threshold down to 0.5~mJy, past the confusion limit. 
This allows detection of objects with $L_{IR} > 5\times10^{11}~L_{\sun}$ out to $z\sim5$ which
transforms to a star-formation rate of $>$85~$M_{\sun}/yr$. On the other hand, the ISO mid-
and far-infrared observations are typically dominated by galaxies at $z<1$ and so can
constrain the low redshift turnover in the luminosity function evolution.

\ifpp
\begin{figure}[ptb]
\plotone{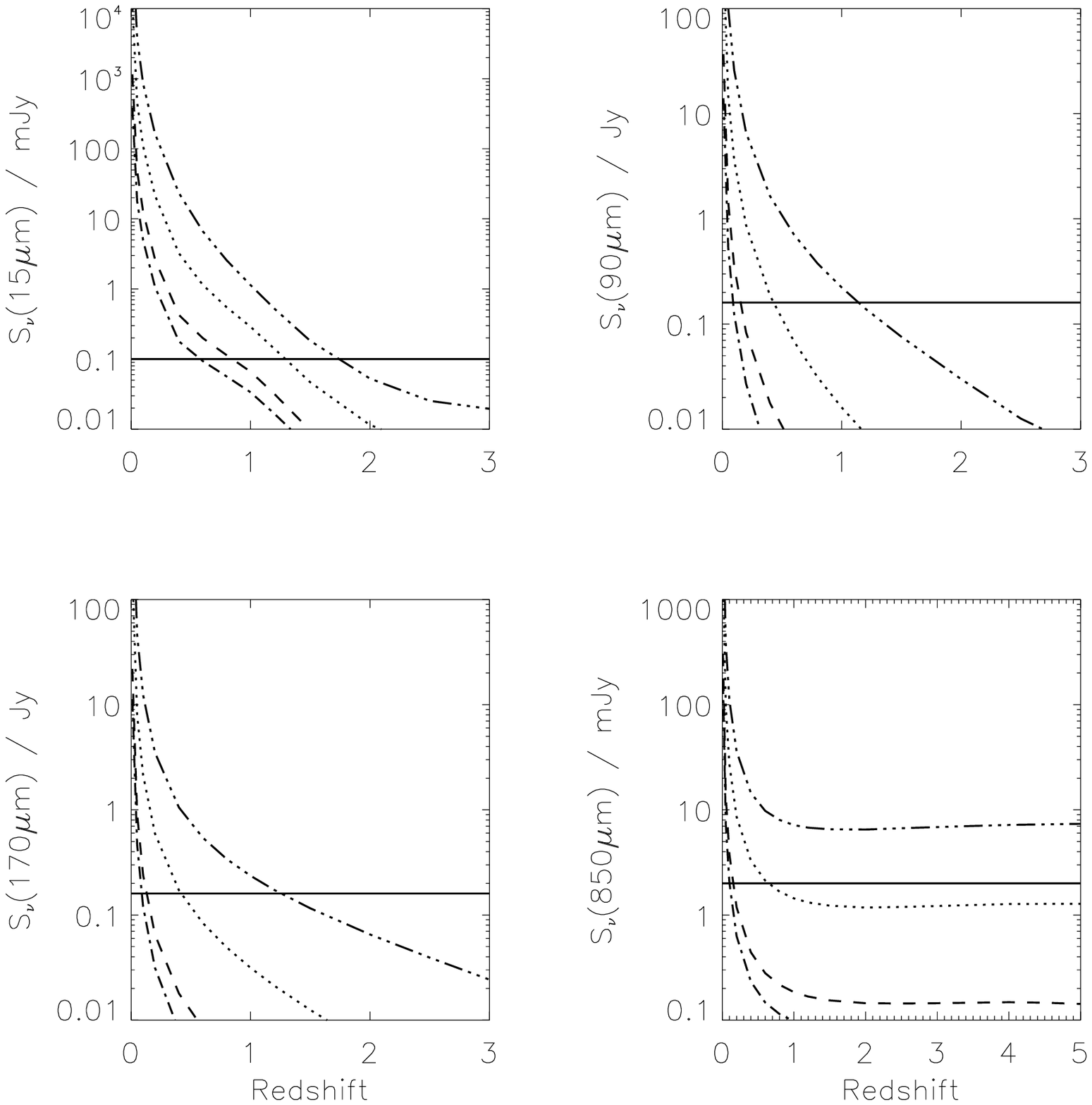}
\caption{Predictions of flux densities at different wavelengths as a function of redshift
for objects with an infrared 
luminosity of $10^{13}$ (dash-dot-dot-dot), $10^{12}$ (dotted), $10^{11}$
(dashed) and 3$\times 10^{10}~L_{\sun}$ (dash-dot). The solid horizontal line is the sensitivity
of the deepest unlensed observation performed. The observations of lensing clusters
at 15~\um\ are sensitive down to 50~$\mu$Jy while those at 850~\um\ are sensitive
down to 0.5 mJy.
}
\label{kcorr}
\end{figure}
\fi

Figure \ref{fig:noevol} illustrates the nature of the counts if
the 15~\um\ luminosity function remained equal to the local one at all redshifts, i.e. no evolution. 
The first plot shows 
the ISOCAM 15~\um\ differential counts
from \cite{Elb99}, which include the IRAS 12~\um\ counts converted to 15~\um\ by \citet{Xu00} and
the ELAIS 15~\um\ counts of
\citet{Ser00} renormalized as in \citet{Gen00}. 
The remaining plots show the
ISOPHOT FIRBACK 170~\um\ counts, ISOPHOT, IRAS BGS and PSCz 90~\um\ counts
and SCUBA 850~\um\ integral counts. Also shown are the relative contributions from
ULIGs, LIGs and $L_{IR} < 10^{11}~L_{\sun}$ galaxies to the counts at different wavelengths\footnote{
From now on we refer to the $L_{IR} < 10^{11}~L_{\sun}$ galaxies as normal/SB galaxies}.
Clearly, some form of redshift evolution in the luminosity function is required to 
fit the counts. It should be emphasized that evolutionary models
should be fit to the differential counts at different wavelengths 
since integral counts tend to smooth over any subtle changes in the galaxy counts slope.
This requires that the calibration of the data from different surveys using
the same instrument be consistent and accurate.

The ISOPHOT 90~\um\ data are known to suffer from large
calibration uncertainties which are not reflected in the error bars.
We show in Section 3.3 that our models fit the 170~\um\ counts but
consistently overpredict the 90~\um\
counts although both wavelengths probe similar populations of galaxies at $z<1.2$. 
We find that an upward correction of the 90~\um\ flux densities 
by 30\%, which is well within the calibration uncertainties, leads to excellent 
agreement between our models and the data.
In addition, the faint end PSCz counts are known to suffer from incompleteness \citep{Efs00}.

\ifpp
\begin{figure}[ptb]
\centerline{\psfig{file=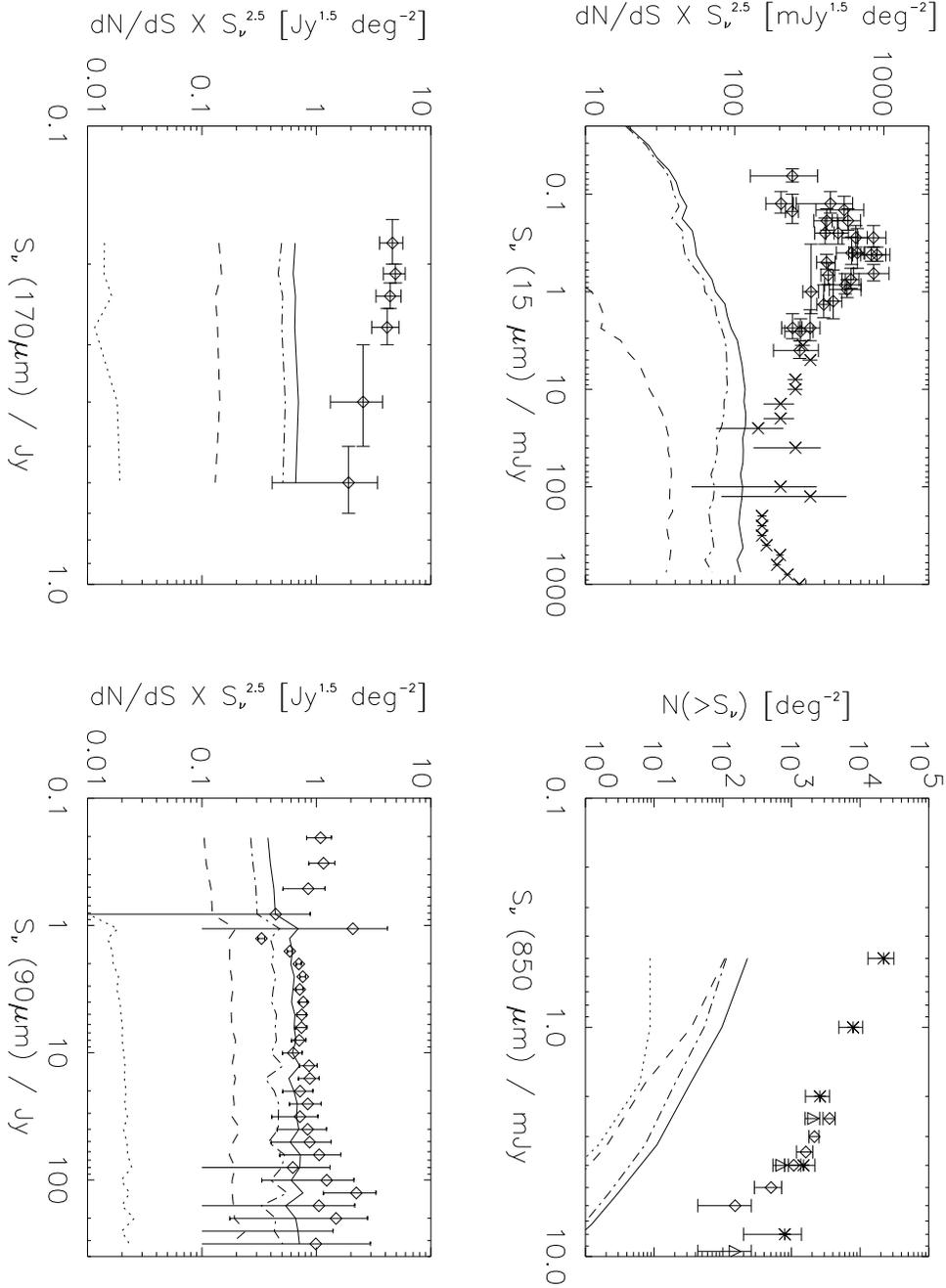,height=7.2in}}
\caption{Results for a no-evolution model with thin solid line representing the contribution
from all galaxies, dotted line indicating ULIGs only, dashed line is LIGs only while dot-dashed line is for normal/starburst galaxies i.e. $L_{IR} < 10^{11}~L_{\sun}$ (This convention will be used
for the remaining figures as well). Clockwise from top left
1. ISOCAM differential
counts at 15~\um. Also shown are the bright end IRAS counts of \citet{Xu00} which potentially have a
factor of 2 uncertainty associated with them, 2.
SCUBA 850~\um\ integral counts, 3. IRAS and ISOPHOT 90~\um\ differential counts 
and 4. ISOPHOT 170~\um\ differential counts (See text for references). 
}
\label{fig:noevol}
\end{figure}
\fi

Any evolution of the luminosity function must have a turnover at 
some redshift $z_{turn}$ to avoid overproducing the CIRB.
In theory, the turnover redshifts $z_{turn}^{D}$ ($z_{turn}^{L}$), 
the $\alpha_{D} (\alpha_{L})$ value for the slope of the density (luminosity) evolution
at $z < z_{turn}$ and the 
$\beta_{D} (\beta_{L})$ value
for $z > z_{turn}$ can be different 
implying that there are effectively six parameters. We however consider models with $z_{turn}^{D}$=
$z_{turn}^{L}$ since it is unclear why the turnover for luminosity and density evolution, 
both of which are probably induced by galaxy interactions,
should be different.
The range of values for $\alpha$, $\beta$ and $z_{turn}$ selected for our models
were $1 \leq \alpha \leq 6$, $-3\leq \beta \leq 0$ and $0.6 \leq z_{turn} \leq 1.5$ respectively. 
Evolutionary models with pure luminosity evolution
and pure density evolution are also considered. 

\subsection{Constraints on Pure Density Evolution}

We find that pure density evolution of the entire luminosity function cannot
reproduce the counts at all the wavelengths. In this scenario, the 15~\um\ counts
are dominated by the normal/SB 
galaxies, not by LIGs
and ULIGs, which is inconsistent with observations in the HDFN+FF. Secondly, the normal/SB 
galaxies are unable to
reproduce the break in the 15~\um\ counts seen at 0.4 mJy but instead produce a sharp break only
at $S_{\nu} < 0.2$~mJy.

However, density evolution models which evolve
just a fraction of the 15~\um\ luminosity function with the fraction being $<$5\%
at $L_{15\mu m}\sim10^{9}~L_{\sun}$ and approaching
100\% at 15~\um\ luminosities greater than 8$\times$10$^{10}~L_{\sun}$ provide
reasonable fits to the data (red broken line in Figure \ref{lbgconn}).
The best fit density evolution parameters then are 
$\alpha_{D}=12.0\pm0.5$ up to
$z_{turn}^{D}=0.7\pm0.1$ followed by $-0.5 < \beta_{D} \leq 0$ (Figure \ref{onlyrho}).
The first plot shows the spectrum of the cosmic infrared background with lower limits
from integrated counts of galaxies in the optical/UV from \citet{Mad00}, measurements
in the near- and far-infrared using the DIRBE
instrument \citep{HAKDO98, Fin00, GWC00, ELW2000}, an estimate of the far-infrared background
from FIRAS \citep{lag99}, and lower limits in the mid-infrared, far-infrared and
submillimeter from counts of individual galaxies \citep{Elb99, Bl99, mat00}. Also
shown is the upper limit on the CIRB from TeV observations of Mrk501 \citep{Sta98} and the
cosmic microwave background at $\lambda > 300$~\um. The remaining plots show the counts
for this evolution model.

Models with $\alpha_{D}>13.0$ result in a significant overprediction of the 170~\um\ counts
while $\alpha_{D}<11.0$ underpredicts the submillimeter counts. Changing the turnover
redshift $z_{turn}^{D}$ to high redshift shifts the knee in the 15~\um\ differential counts to
fainter flux levels and vice-versa. The slope of the evolution at $z_{turn}^{D} > 0.7$
is mainly constrained by the spectrum of the CIRB at $\lambda > 200$~\um\ and the 
850~\um\ counts. 

The 170~\um, 90~\um\ and 15~\um\ observations all probe the population of galaxies at $z<1.2$.
However, in this density evolution model, the 15~\um\ counts are dominated ($\sim$90\%) by LIGs
and have $\sim$5\% contribution from ULIGs.
In comparison, the 170~\um\ counts are dominated by ULIGs at redshifts between 0.5$-$1 while
the 90~\um\ counts have roughly equal contributions from all three populations of galaxies.
The 850~\um\ galaxies at flux densities larger than 1 mJy are mainly ULIGs 
at $z > 1$ while LIGs between redshifts of $0.6-2.0$ dominate the counts at fainter flux levels.

The pure density evolution model shown appears to overpredict the
contribution from LIGs ($\sim$90\%) to the 
15~\um\ number counts and underestimates the bright end 850~\um\ counts. 
Secondly, many LIGs and ULIGs have been 
morphologically associated with disturbed systems. So a steep evolution in the density
of objects should reflect in an increase of the merger fraction which is defined as the
fraction of galaxies in close pairs.  Observationally,
the merger fraction when averaged over all galaxies appears to evolve much slower with redshift,
approximately as $(1+z)^{3}$ \citep[e.g.][]{Lef00}. It is possible though
that the LIGs and ULIGs have a merger fraction that increases much more rapidly than
$(1+z)^{3}$ but this has not been estimated since there are no clear
observational signatures of LIGs and
ULIGs at visible wavelengths. Although there is no strong observational evidence
in favor of pure density evolution of a fraction of the luminosity function, 
we show in Figure \ref{lbgconn} that it does
predict the same number density of infrared luminous galaxies at high redshift 
as other models and so cannot be entirely ruled out.

\ifpp
\begin{figure}[ptb]
\centerline{\psfig{file=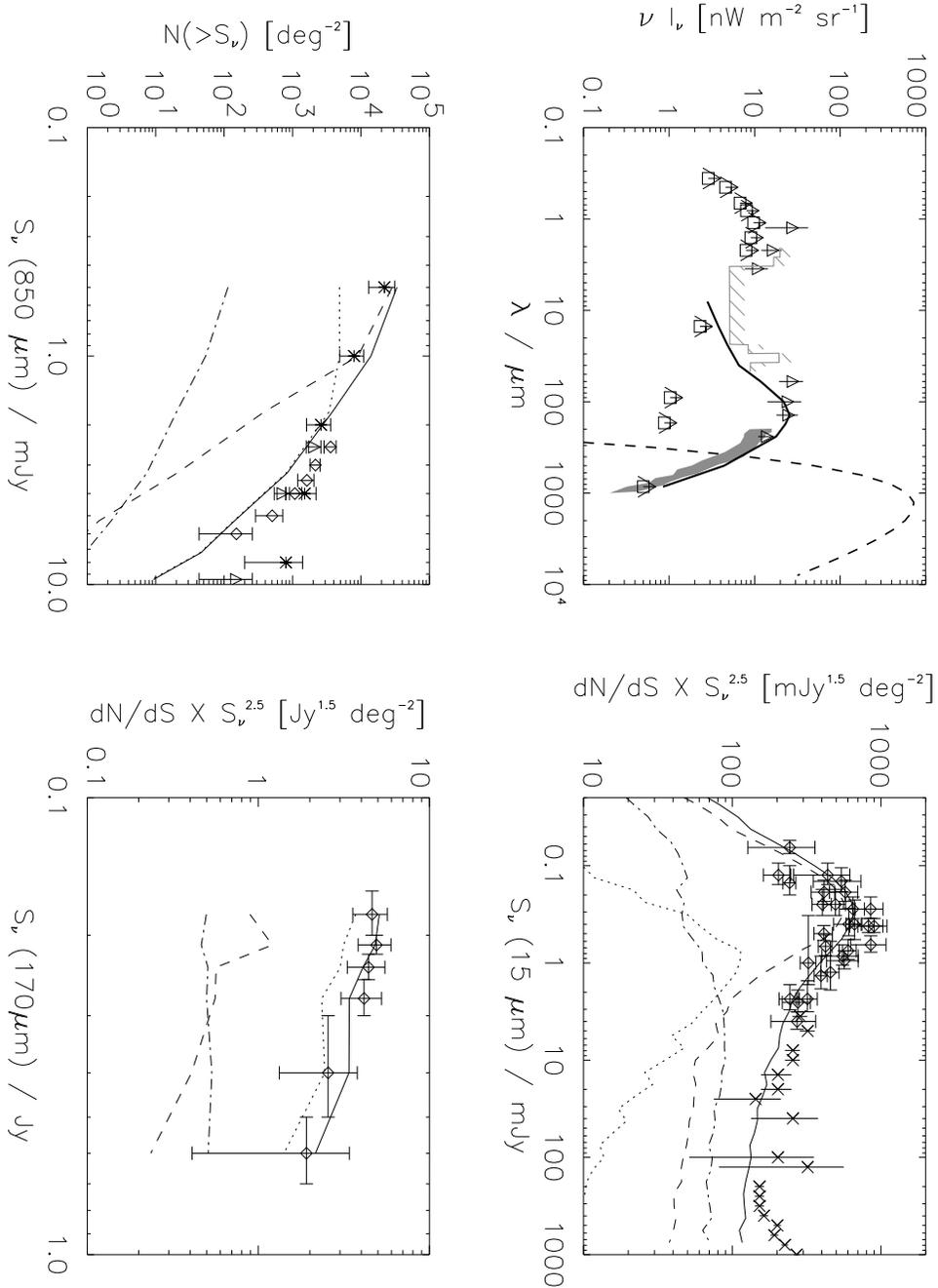,height=7.2in}}
\caption{Results for the best fitting pure density evolution model with only the most
luminous end of the LLF evolving (Red dashed line in Figure \ref{lbgconn}). 
The evolutionary parameters are $\alpha_{D}=12.0$,
$\beta_{D}=0$, $z_{turn}^{D}=0.7$. The upper left plot shows the spectrum of the
cosmic infrared background: triangles
represent the DIRBE/COBE results, squares with upward arrows are lower limits from integrated
counts at different wavelengths, the dark band is the FIRAS constraint, hatched region represents
upper limits from TeV gamma ray observations of Mrk501, dashed line represents the cosmic
microwave background, thick solid line is the prediction from the model. The remaining plots
are similar to Figure \ref{fig:noevol}.
}
\label{onlyrho}
\end{figure}
\fi

\subsection{Constraints on Pure Luminosity Evolution}

We have shown above
that some form of luminosity evolution is required to avoid deriving
large values for the slope of the density evolution. 
The slope of the luminosity evolution $\alpha_{L}$ is strongly constrained by the
mid-infrared number counts. Values of $\alpha_{L} < 4.5$ are unable to reproduce the break of the
differential counts seen at a 15~\um\ flux density of 0.4~mJy while $\alpha_{L} > 5.5$ results
in an overproduction of the counts at bright flux densities (Figure \ref{onlylum}).
If the entire LLF is evolved (solid black line in Figure \ref{lbgconn}), 
$\alpha_{L} = 5.0$ up to $z_{turn}^{L}$=0.8
followed by $-0.5 < \beta_{L} \leq 0.0$, 
the 90~\um\ differential counts are overproduced but all the other counts are reproduced very well.

This is only partially consistent with the results of \citet{Xu00} who suggested that
the mid-infrared counts can be modelled by evolving the entire LLF by
$L(z)\propto (1+z)^{4.5}$ for $z<1.5$
and by $L(z)=L(0)\times 2.5^{4.5}$ for higher redshifts. 
We find
that a luminosity evolution of $(1+z)^{4.5}$ up to $z\sim1$ followed by $L(z)=L(0)\times 2.0^{4.5}$
overpredicts the CIRB and provides only marginal fits to the
mid-infrared counts at the faint end. Extending this evolution up to $z_{turn}^{L}\sim1.5$
severely overpredicts the CIRB as well as
the observed faint end 15~\um\ and 850~\um\ counts. 

The main problem with pure luminosity evolution is that the 90~\um\ counts
at $S_{\nu} < 6$~Jy are overproduced but as mentioned earlier,
this can be resolved by rescaling the ISOPHOT 90~\um\ flux densities
upwards by 30\% which is within the calibration uncertainties of the instrument.
In addition, the break in the 15~\um\ differential counts in this evolution
model is not as sharp as observed.

As in the pure density evolution model, the counts 
at 15~\um, 90~\um\ and 170~\um\ all trace galaxies at $z<1.2$ but
the relative contributions from LIGs, ULIGs and normal/SB galaxies differ, with the 90~\um\
counts having roughly equal contributions from all three populations, the 170~\um\ counts being
dominated by ULIGs and the 15~\um\ counts being dominated by LIGs and low redshift normal/SB galaxies.
Pure luminosity evolution predicts that the contribution from normal/SB galaxies
to the 15~\um\ counts between 0.1 and 0.4 mJy is 35\%, similar to that observed 
in the HDFN+FF. 
However, the contribution from ULIGs in the same flux density range is found to be
only 6\% in this model; a factor of $3-4$ smaller than that observed.
 
\ifpp
\begin{figure}[ptb]
\centerline{\psfig{file=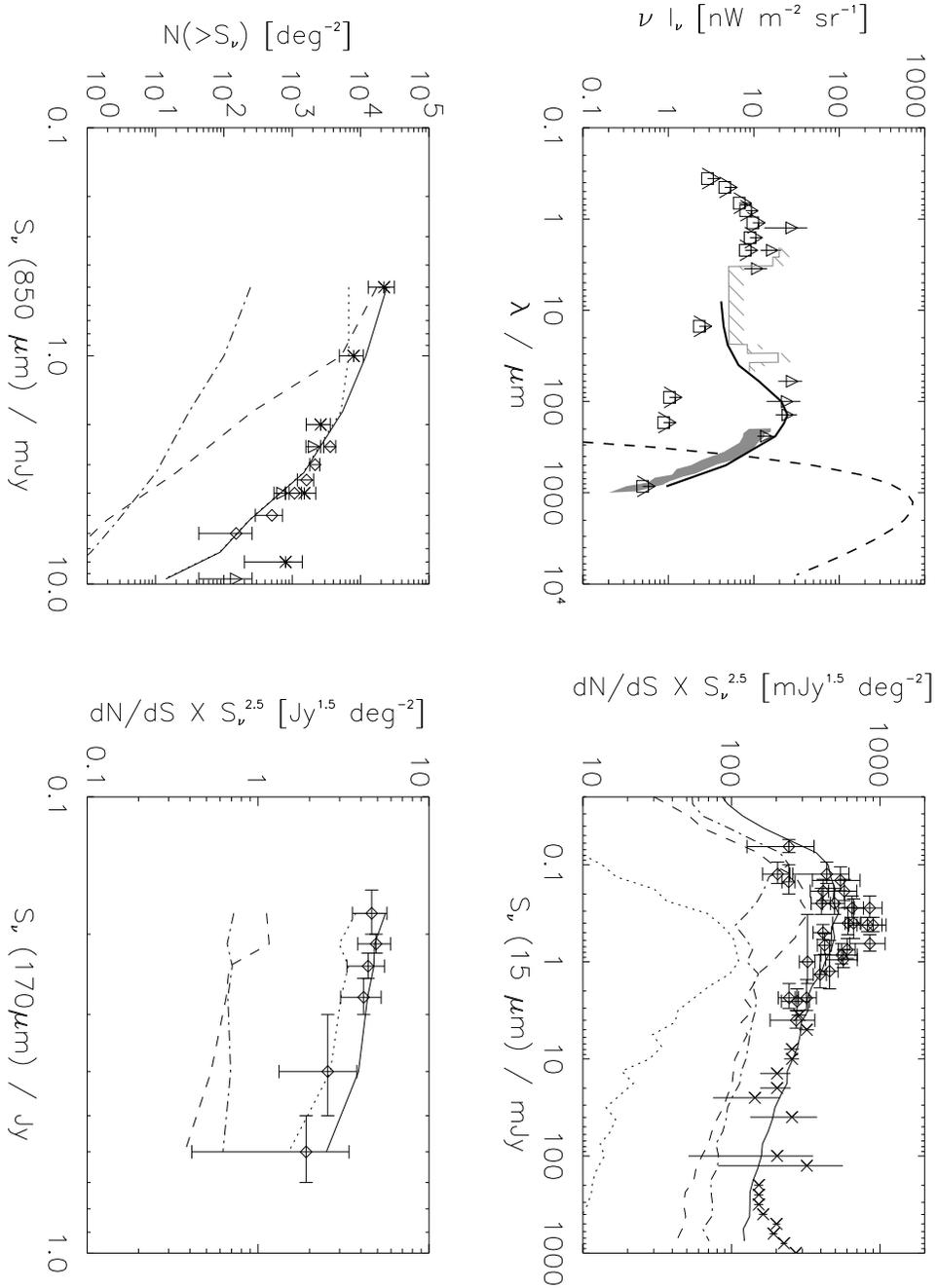,height=7.2in}}
\caption{Results for the best fitting pure luminosity evolution model of the entire
LLF (Solid black line in Figure \ref{lbgconn}) with $\alpha_{L}=5.0$,
$\beta_{L}=0$, $z_{turn}^{L}=0.8$. $\beta_{L}=-0.25$ reduces the CIRB at $\lambda > 200$~\um\ 
but slightly lowers the bright end SCUBA counts at flux densities greater than 2 mJy. 
}
\label{onlylum}
\end{figure}
\fi

\subsection{Combination of Luminosity and Density Evolution}

We have already illustrated the degeneracy and problems with pure luminosity and pure density
evolution in the fits to the number counts. As illustrated above, both of them
provide reasonable fits to the spectrum of the CIRB and to the differential counts
at three wavelengths. An additional degeneracy is introduced when using a combination
of luminosity and density evolution with only a fraction of the LLF evolving.

\ifpp
\begin{figure}[ptb]
\centerline{\psfig{file=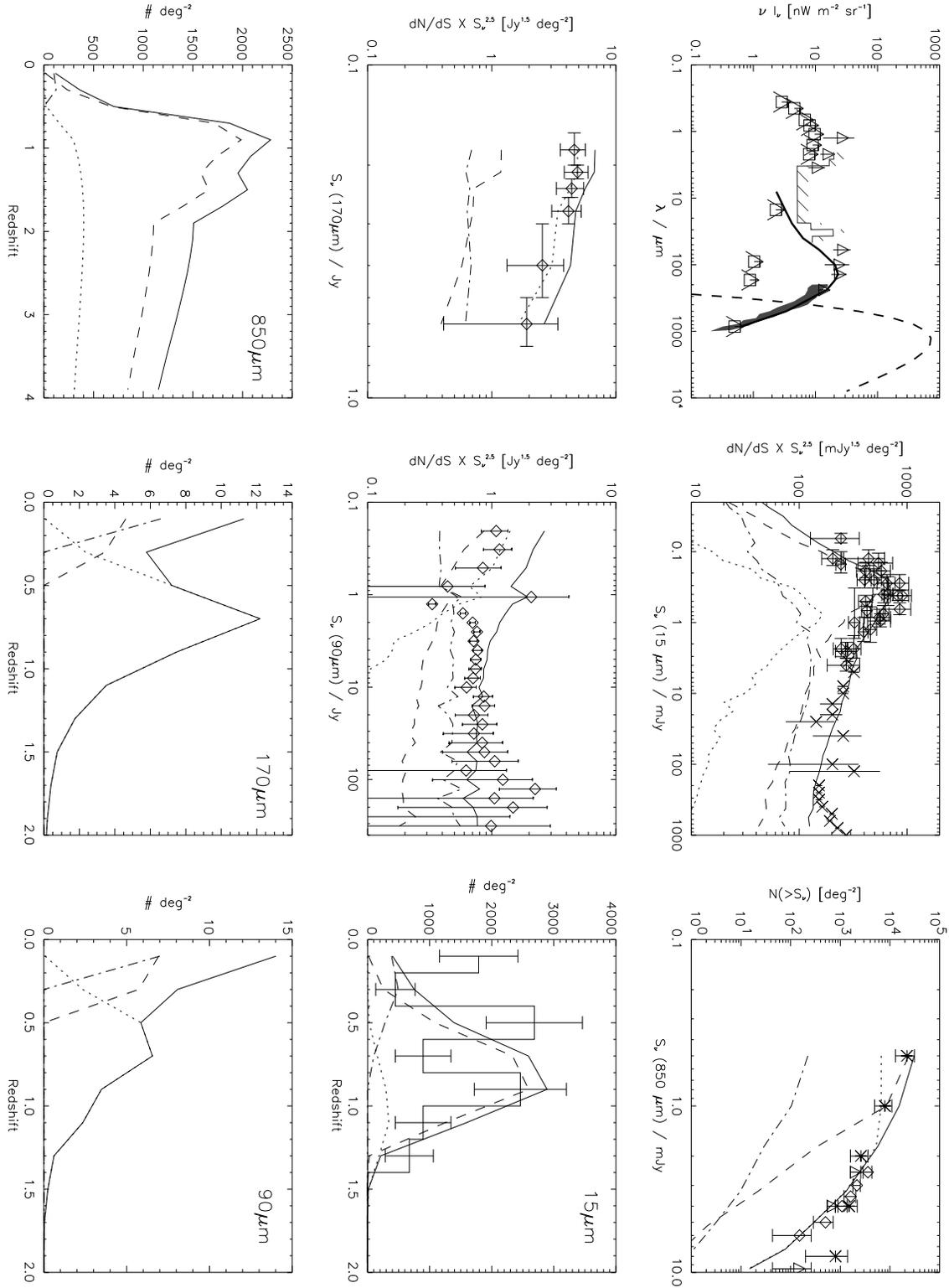,height=8.2in}}
\caption{Results for a model with both luminosity and density evolution that evolves 5\%
of the luminosity function at $L_{IR} < 10^{10.2}~L_{\sun}$ and 100\% at
higher luminosities (purple broken line in Figure \ref{lbgconn}). 
$\alpha_{L}=4.5$, $\alpha_{D}$=1.5, $\beta_{L}=0$, $\beta_{D}=-0.4$,
$z_{turn}^{L}=z_{turn}^{D}=0.8$.
}
\label{denslum}
\end{figure}
\fi

The fraction of the LLF that is evolving is defined as the ``dusty starburst''
population. 
We consider these to be galaxies with $L_{B}/L_{IR} < 0.5$. 
This corresponds to $L_{IR} \geq 10^{10.2}~L_{\sun}$. 
All galaxies with $L_{IR} \geq 10^{10.2}~L_{\sun}$ are then evolved while only 
about 5\% of the galaxies with $L_{IR} < 10^{10.2}~L_{\sun}$ (``normal'' galaxy
population) is evolved, with a smooth transition between the 
two (purple broken line in Figure \ref{lbgconn}). 
We also considered
models where the dusty starburst population is defined at a larger minimum
luminosity ($L_{IR}\sim10^{11}~L_{\sun}$) but were unable to find evolution parameters
that could reasonably reproduce all the data.

Our best fit model using a combination of both density and luminosity evolution
for a fraction of the LLF as defined above is shown in Figure \ref{denslum}. Almost all the counts
are reproduced quite well, with the exception of the 90~\um\ counts from
ISOPHOT and PSCz which all our models consistently over estimate. 
We interpret this to be due to a calibration error in the ISOPHOT data as a result
of which the published flux density values are low by $\sim$30\%.
Also shown in the figure is
the surface density of galaxies per redshift bin 
contributing to the counts at different
wavelengths, as derived from the model, and the observed redshift 
distribution of the ISOCAM 15~\um\ galaxies in the HDFN+FF \citep{Auss01}. The 
size of each redshift bin is 0.2.

In the preceding three subsections, we have shown a range of models that
evolve the local 15~\um\ luminosity function and
fit the observed counts at mid- and far-infrared wavelengths as well as the spectrum
of the CIRB.
Ultradeep SIRTF observations at 24~\um\ can potentially break the degeneracy
in these models
if the counts can be determined to an accuracy of 20\% or better (Figure \ref{srtf}). 
The range of integral source
counts that we predict based on our three evolutionary
models (density, luminosity and combination of both)
are 4.7, 3.8 and 3.7 per $\sq\arcmin$
for $S_{\nu} > 120~\mu$Jy and 19.8, 19.3 and 11.8 per $\sq\arcmin$ for $S_{\nu} > 22~\mu$Jy
respectively. 
However,
the integral counts can be as low as 9.1/$\sq\arcmin$ at a flux density limit of 22~$\mu$Jy
for models which are at the extreme lower limit of the uncertainty in current observations.

\ifpp
\begin{figure}[ptb]
\plotone{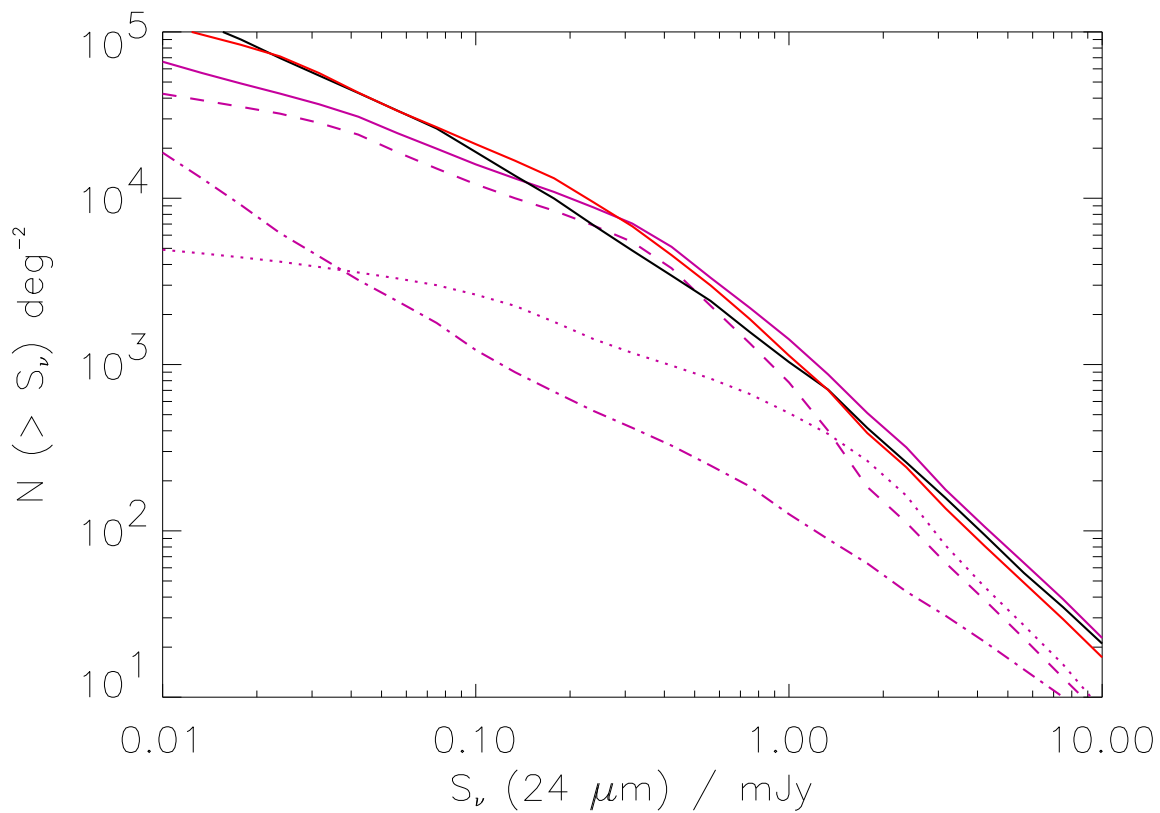}
\caption{
Prediction for integral source counts seen by SIRTF at 24~\um\ for the three different evolutionary
models described in the text. Red line is for pure density evolution, black line
is for pure luminosity evolution, solid purple line is for density+luminosity evolution
while the broken lines are contributions from ULIGs, LIGs and $L_{IR}<10^{11}~L_{\sun}$ galaxies
to the counts from the density+luminosity evolution using the 
convention defined in Figure \ref{fig:noevol}.
}
\label{srtf}
\end{figure}
\fi

\section{The Origin of the CIRB}

\subsection{Nature of the Galaxies Contributing to the CIRB}

In our evolution models, we have assumed that the contribution from AGN to the
counts and the cosmic background is insignificant. 
Other evolutionary models, which assumed
an AGN component, arrived at the same conclusion \citep{Mal98, Row00, Xu00b, Fra00}.
Observational evidence for this assumption
comes from deep {\it Chandra} observations of the HDFN proper \citep{Brandt01}
which detected 8 of the ISOCAM 15~\um\ sources.
However, only one of these is an
AGN at $z\sim$1 and this was already known as such from 
observations at visible wavelengths \citep[see discussion in][]{Elb00, Auss01}.
It should be noted that it is insufficient to have an AGN in a galaxy
to violate this assumption but that the integrated infrared light of the galaxy must
be dominated by an AGN rather than by star formation.
However, since the contribution of dust obscured AGN at high redshift, beyond ISOCAM
detection thresholds, is unknown, our results are 
subject to this uncertainty. A large ($>$ 20\%) contribution from AGN to 
the source counts or the CIRB will imply a weaker redshift evolution of the luminosity function.

The evolution parameters in our model are constrained strongly by the ISOCAM counts
at $z<1.2$ and by the SCUBA counts at $z\sim1-3$. In addition, the ISOCAM counts are dominated
by LIGs while the SCUBA counts are dominated by ULIGs. Thus, barring a dramatic change
in the ratio of LIGs to ULIGs between a redshift of 1 and 2, we conclude that our
models have robustly determined the evolution of the luminous end ($L_{IR} > 10^{11}~L_{\sun}$)
of the LLF up to $z \sim 2$.
At $z >> 2$, the best constraint comes from the spectrum of the CIRB. Since all our
models which are almost flat beyond $z\sim2$ provide values for the CIRB that
are at the upper limit of the values observed by FIRAS, we conclude that these models
place a strong upper limit on the estimate of dust enshrouded star formation at $z > 2$.

\begin{deluxetable}{cccc}
\tablecaption{Origin of the CIRB}
\tablehead{
\colhead{Wavelength} & \multicolumn{2}{c}{$\nu I_{\nu}$} & \colhead{Contribution from $z<1.5$ galaxies}  \\
\colhead{$\mu m$} & \multicolumn{2}{c}{nW m$^{-2}$ sr$^{-1}$} & \colhead{} \\
\cline{2-3}
\colhead{} & \colhead{Observed} & \colhead{Model} & \colhead{}
}
\startdata
15~\um & 2.4$\pm$0.5 & 3.2 & 90\% \\
24~\um & ... & 4.2 & 83\% \\
140~\um & 25$\pm$7 & 23.1 & 82\% \\
240~\um & 14$\pm$3 & 15.1 & 67\% \\
850~\um & 0.5$\pm$0.2 & 0.63 & 28\% \\
\enddata
\label{tab1}
\end{deluxetable}

\begin{deluxetable}{cc}
\tablecaption{Origin of the 140~\um\ EBL as derived from the models}
\tablehead{
\colhead{Source Type} & \colhead{Contribution} 
}
\startdata
ULIGs & 25\% \\
LIGs & 60\% \\
$L_{IR}<10^{11}~L_{\sun}$ galaxies & 15\% \\
 & \\
ISOCAM galaxies with $S_{\nu}(15~\mu m) > 0.1$~mJy & 63\% \\
ISOCAM galaxies with $S_{\nu}(15~\mu m) > 0.05$~mJy & 73\% \\
 & \\
SCUBA galaxies with $S_{\nu}(850~\mu m) > 2$~mJy & 15\% \\
SCUBA galaxies with $S_{\nu}(850~\mu m) > 0.5$~mJy & 71\% \\
\enddata
\label{tab2}
\end{deluxetable}

Our models indicate that 
about 80\% of the 140~\um\ CIRB is produced at $z < 1.5$. In comparison, 90\% of the 15~\um\
EBL, 65\% of the 240~\um\ background
and only about 30\% of the 850~\um\ background, is produced within this redshift 
range (Table \ref{tab1}).
We also derive that ISOCAM galaxies brighter than 0.1 mJy contribute 14.5 nW~m$^{-2}$~sr$^{-1}$ to
the 140~\um\ EBL, while galaxies brighter than 0.05 mJy produce 16.8 nW~m$^{-2}$~sr$^{-1}$.
This accounts for about 75\% of the total far-infrared background (Table \ref{tab2}). In comparison,
we find that the contribution from SCUBA detected galaxies brighter than the confusion limit of
2 mJy at 850~\um\ is only
3.4 nW~m$^{-2}$~sr$^{-1}$ at 140~\um\ while galaxies brighter than 0.5 mJy produce
16.4 nW~m$^{-2}$~sr$^{-1}$. 
This is because at 850~\um\ flux densities fainter than 1~mJy, the LIGs which dominate
the ISOCAM counts and produce the majority of the 140~\um\ EBL contribute 
significantly to the SCUBA counts.
The total EBL
at 15~\um\ from our model is 3.2 nW~m$^{-2}$~sr$^{-1}$. 
The EBL obtained by integrating the observed ISOCAM counts above 50~$\mu$Jy
is 2.4$\pm$0.5 nW~m$^{-2}$~sr$^{-1}$ \citep{Elb00}, hence as much
as 73$\pm$15\% of the 15~\um\ background might have already been resolved by ISOCAM.
The models indicate that ULIGs contribute 15\% of the 15~\um\ EBL observed by ISOCAM above 0.1 mJy, 
LIGs about 65\% and normal and low luminosity
starburst galaxies the balance. 
In comparison, at 140~\um, we find that ULIGs contribute 25\% of the CIRB,
LIGS contribute 60\% and normal/SB galaxies the balance (Figure \ref{cirbcont}). Thus, 
infrared luminous galaxies which appear to be indistinguishable from normal
galaxies in terms of their optical/near-infrared
luminosity and form a negligible part of the energy budget in the local
universe, dominate the star formation and therefore the energy budget at redshifts $z\sim1-3$.

\ifpp
\begin{figure}[ptb]
\plotone{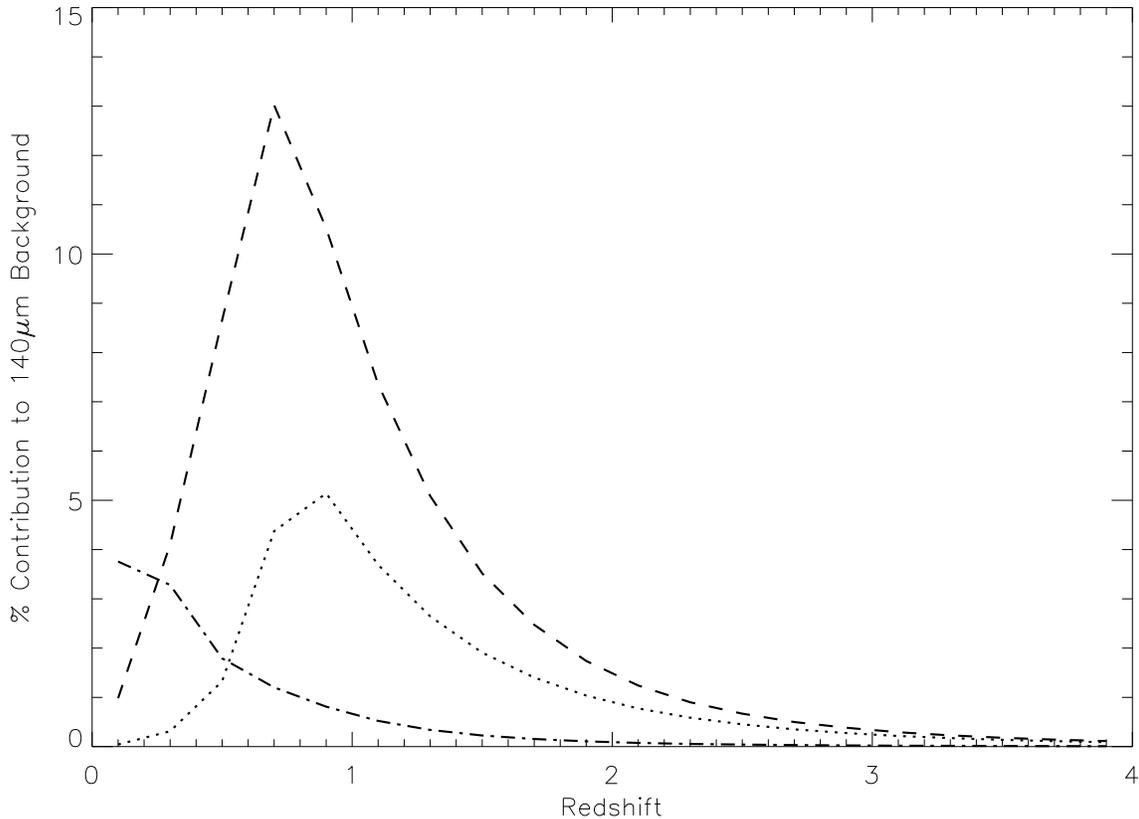}
\caption{
Plot showing the relative contribution of LIGs (dash), ULIGs (dots)
and normal/starburst galaxies (dot-dash)
to the 140~\um\ extragalactic background light as a function of redshift. Each redshift bin is 0.2.
}
\label{cirbcont}
\end{figure}
\fi

\subsection{Infrared Luminosity Function and the LBG Connection}

The first panel of Figure \ref{lbgconn} shows the 15~\um\ local luminosity function (LLF)
along with the fraction of galaxies that are evolved in the different models.
For the density evolution model, the broken red line is evolved. For the luminosity
evolution model, the whole 15~\um\ LLF shown as the solid black line, is evolved.
For the model with density+luminosity evolution, the broken purple line
is evolved. In the pure density evolution and density+luminosity evolution model, there 
is a non-evolving component with a constant
comoving density which corresponds to the difference between the total LLF and
the evolving component.

Although the fraction of the luminosity
function that is evolving and the evolutionary parameters are significantly different
in our three evolutionary scenarios, we find that our models predict similar
comoving number densities of infrared luminous galaxies at high redshift. This is illustrated
in two panels of Figure \ref{lbgconn} which show the derived 15~\um\ luminosity function
at redshifts of 0.4 and 0.8. The models also provide
strong evidence for a change in the shape of the infrared luminosity function.
The comoving number density of luminous infrared galaxies has to increase by more than
two orders of magnitude between redshifts of 0 and 1, to fit the ISOCAM and SCUBA counts. 
The faint end of the LLF cannot be enhanced by the same factor since this would lead to 
an over-production of the CIRB although these galaxies would be below the sensitivity limit
of the long wavelength surveys. Lastly, as much as 85\% of the far-infrared background
can be attributed to infrared luminous galaxies. This implies that the contribution from
normal and low-luminosity starburst galaxies ($L_{IR}<10^{11}~L_{\sun}$) to the dust
enshrouded star formation rate is relatively small. 
So, estimates of the total SFR by applying
a constant extinction correction to all optical/UV selected galaxies are incorrect.
We conclude that 
long-wavelength surveys between 15~\um\ and 850~\um\ that probe 
galaxies at the luminous end of the infrared luminosity
function provide a very effective way of tracing the bulk of the dust obscured star formation.

The connection between infrared luminous galaxies and the Lyman-break galaxy (LBG) population
is intriguing. Figure \ref{lbgconn} shows a 
comparison between the LBG 60~\um\ luminosity function 
at $z\sim3$ of \citet[][hereafter AS00]{Adel00}, which was 
derived based on an extinction correction to
optical/UV data as a function of the UV slope of individual galaxies, and 
our equivalent 60~\um\ luminosity function at
$z\sim3$, which we have argued earlier is only a strong upper limit.
The agreement
is extremely good considering that they were estimated in completely independent ways. 
The AS00 luminosity function predicts almost the same luminosity function as our
estimate from the pure luminosity evolution model to within 50\%. It is discrepant
with the luminosity functions from our other two models by as much as an order of magnitude 
at the faint end but only a factor of 2 at the bright end.
As mentioned earlier, the long wavelength surveys which constrain our models
are mainly sensitive to the evolution of galaxies at the bright
end of the luminosity function at $z < 2$. We are unable to constrain with much certainty
the evolution of the faint end of the luminosity function although we do
place an upper limit based on the observed intensity of the CIRB. 
Furthermore, our estimates at $z\sim3$ are only a strong upper limit to the
number density of infrared luminous galaxies since any
further evolution at high redshift overproduces the CIRB at $\lambda > 200$~\um\ while a 
decay in the evolution at $z>2$ as $[1+(z-2)]^{-2}$ is marginally consistent with both
the submillimeter counts and the CIRB spectrum. 
Thus, we conclude that 
optical/UV surveys that trace the LBG population at $z > 3$,
after an extinction correction factor that spans the range 2$-$100, provide a 
good estimate of 
dust obscured star-formation at high redshift. They complement the results of future
mid- and far-infrared surveys with SIRTF which will be able to directly
observe the dust emission of LBGs with LIG-type infrared luminosities up to $z\sim2.5$.

There is observational evidence for the LBG population to be distinct from
the bright ($S_{\nu}(850~\um) > 6$~mJy) submillimeter galaxies \citep{bar00, chap00b}. This is 
because most of the bright 850~\um\
galaxies are extreme ULIGs with $L_{IR} > 10^{12.6}~L_{\sun}$. The AS00 observations
detect only 2 out of 831 galaxies above this detection threshold and only 27
of their LBG sample would have a submillimeter detection above the level of $\sim$1~mJy.
Secondly, at $z\sim3$, 850~\um\ observations would probe 
rest frame $\sim$200~\um\ emission.
For a given far-infrared luminosity, the 60~\um\ luminosity shows
a factor of 2$-$3 less scatter than the 850~\um\ luminosity among local galaxies 
(Figure \ref{fig:compcorr}) suggesting that the $\lambda > 200$~\um\ spectral shape
of galaxies might potentially have a larger scatter which 
would lead to uncertain flux estimates on a galaxy
by galaxy basis. 
Lastly, we do not find any
scenario where the contribution from ULIGs is greater than 30\% of the comoving
star formation rate at $z\sim3$. Naturally, the contribution from extreme ULIGs
traced by the bright submillimeter galaxies is even smaller. 
Thus, although the contribution to the star formation rate density from extreme ULIGs is missed
in observations of the LBG population, their contribution is significant only at the
level of $<$10\% and hence they are less important to an estimate of the high-redshift
dust obscured star formation.

\ifpp
\begin{figure}[ptb]
\plotone{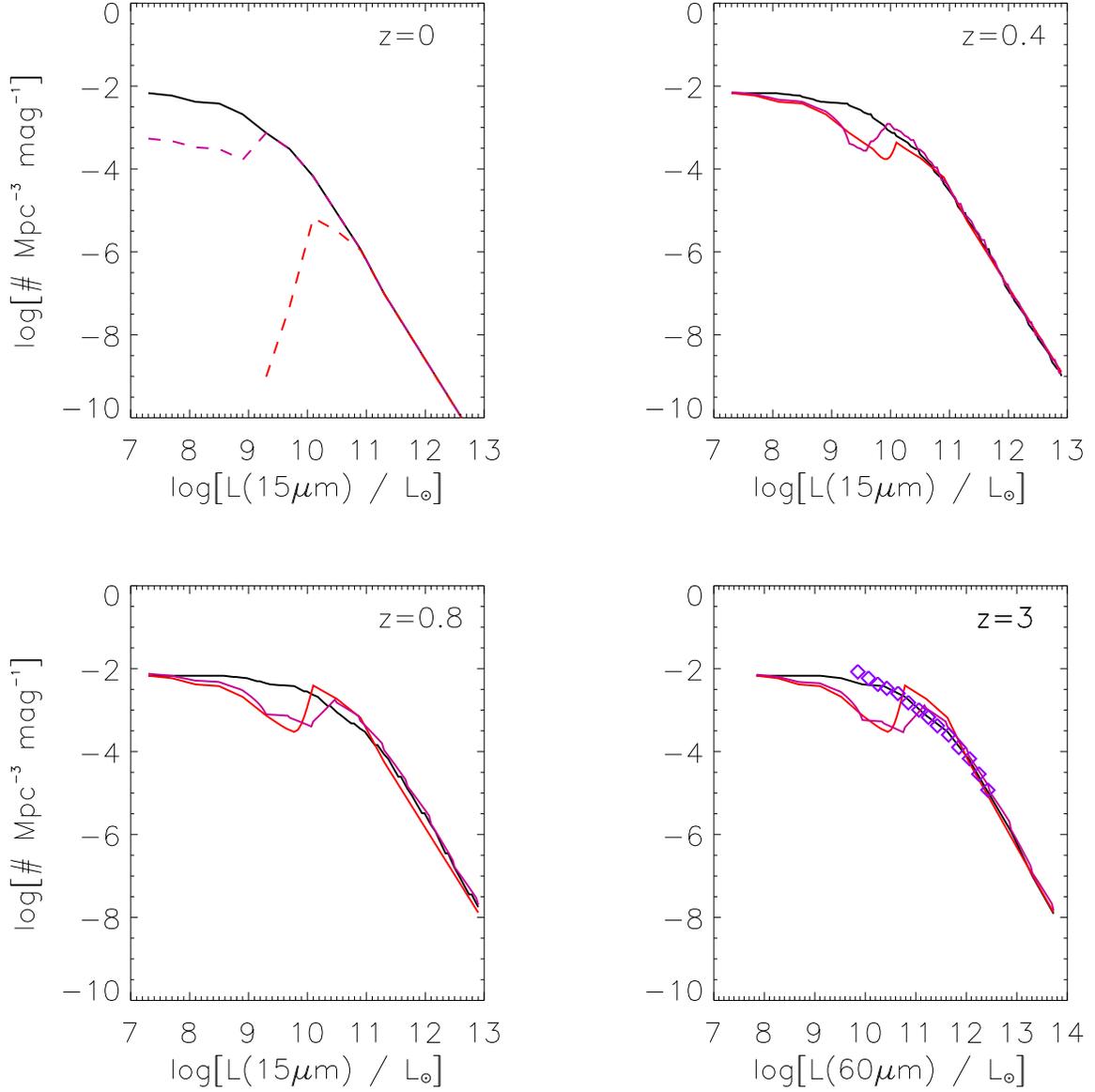}
\caption{Top left: The local 15~\um\ luminosity function (LLF; solid line).
For pure luminosity evolution, we evolved the whole LLF, the purple dashed line is the
starburst component which is evolved in the density+luminosity evolution model, the red dashed
line is the luminous component that is evolved in pure density evolution models.
Top right: Total (evolving+non-evolving) 15~\um\ luminosity function for 
the three models at $z=0.4$, 
Bottom Left: Total 15~\um\ luminosity function at $z=0.8$,
Bottom Right: 15~\um\ luminosity
function at $z=3$ converted to 60~\um\ and compared with the 60~\um\ LBG luminosity function
(purple diamonds) of \citet{Adel00}.
}
\label{lbgconn}
\end{figure}
\fi

\subsection{The Revised Star Formation History of the Universe}

Having constrained the evolution of the mid-infrared and thereby the far-infrared 
luminosity functions, we can
derive the evolution of the dust enshrouded star formation rate with redshift.
Using the equations listed in Section 2, our derived comoving SFR from all galaxies
is shown in
Figure \ref{sfrhis} while the separate contribution from LIGs and ULIGs is shown in
Figure \ref{sfrhis2}. Figure \ref{sfrhis} also shows the absolute minimum and maximum range of 
dust obscured star formation rate values. The maximum values are derived from models which 
marginally overproduce the CIRB and the counts. The minimum values are derived by
using an evolutionary model which is marginally consistent with the observations (Section 4.2) and
by only considering the contribution from LIGs and ULIGs since those are the only
galaxies which are directly observed at high redshift in the 15~\um\ and 850~\um\ surveys.
Also shown is the SFR as inferred
from direct observations at visible/UV/near-infrared
wavelengths \citep{Lil96, Mad96, Con97, Cow99, St99, Yan99}, SFR estimates obtained from 
extinction corrections to these observations \citep{Mad98, Meu99, St99, Tho00}, and 
SFR derived from ISOCAM observations of the Canada France Redshift Survey (CFRS)
field \citep{Flo99}. In addition, lower limits to the SFR from radio measurements and two points
representing the completeness corrected submillimeter observations are also shown
\citep{bar00}. The 1$\sigma$ uncertainty in our derived rate is about 50\%
and is primarily dependent on the transformation from 15~\um\ to infrared luminosities 
which as derived earlier has a 1$\sigma$ of 40\%, and the transformation from infrared luminosities
to star formation rate which assumes a Salpeter initial mass function (IMF)
and also has an uncertainty of about 30\% \citep{Kenn98}.

The dust enshrouded SFR density peaks at a 
redshift of 0.8$\pm$0.1 with a 
value of 0.25$^{+0.12}_{-0.1} M_{\sun} yr^{-1} Mpc^{-3}$. 
In a (0.3,0.7,75) cosmology this corresponds to 6.2 Gyr after the Big Bang.
The dusty SFR then remains almost constant up to $z\sim 2$ which corresponds to an age
of 3 Gyr beyond which this value provides a strong upper limit to the amount of dust obscuration. 
This is similar to the shape of the star formation history preferred by \citet{Sad01} in their analysis of the CIRB.
Our values are a factor of 2 larger than 
estimates at $z\sim1$ from $H\alpha$ observations by
\citet{Yan99} and a factor of 3-7 larger than extinction uncorrected optical/UV observations at 
$z < 2$. The models are in excellent agreement with the submillimeter
data corrected for incompleteness \citep{bar00} but are higher than the \citet{St99} extinction
corrected points. 
The values we derive are systematically
higher than those in \citet{Gis00} but within the uncertainties especially if
the difference in the $L_{IR}$-to-SFR calibration coefficient is factored in. 
Our models also indicate a faster evolution at $z<1$ than the models
of \citet{Bl99b} 
which is not surprising since they did not use the ISOCAM data to constrain their low redshift
evolution. However, our high redshift plateau is similar to their `Anvil-10' model.

\ifpp
\begin{figure}[ptb]
\centerline{\psfig{file=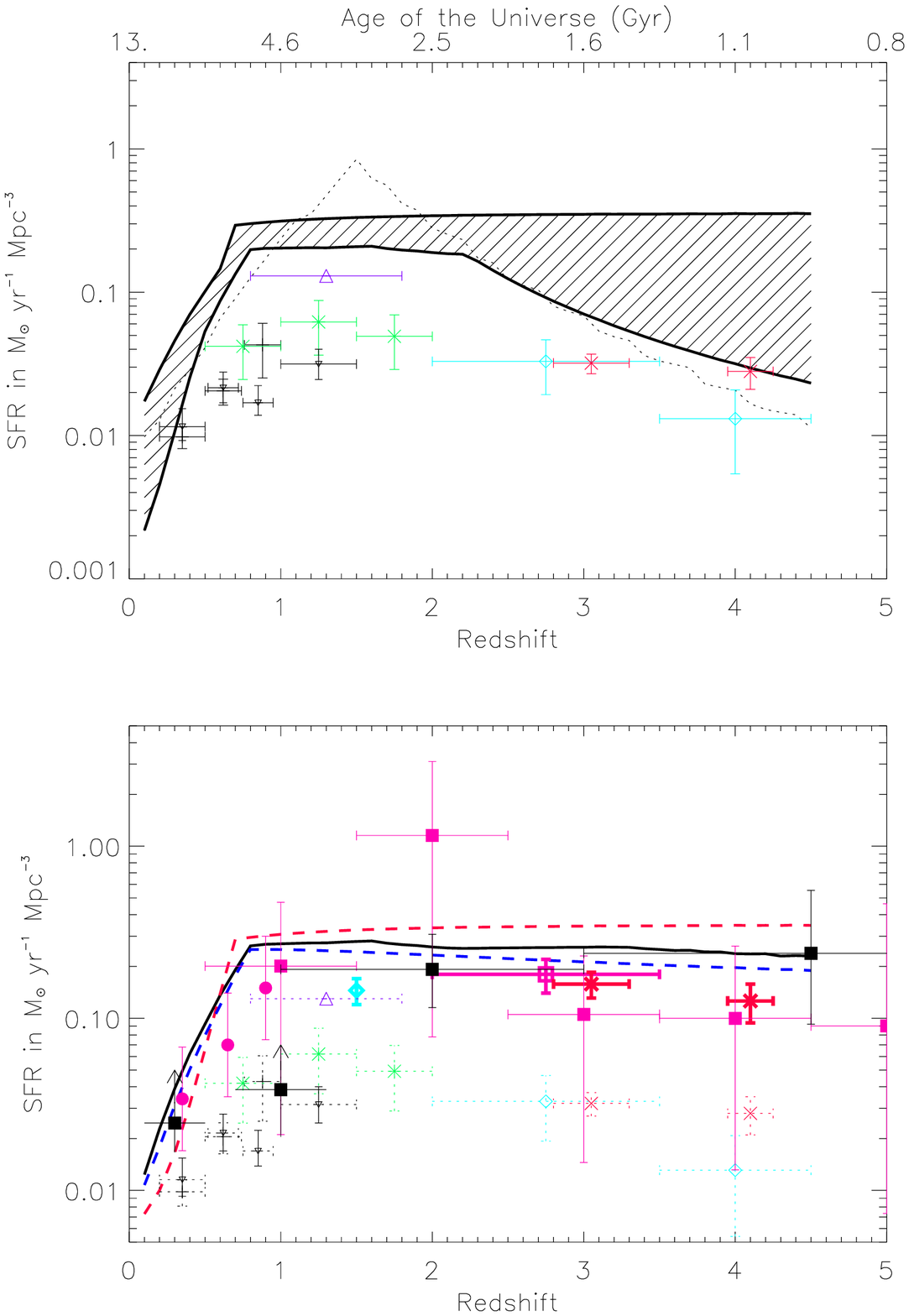,height=5.0in}}
\caption{
The upper plot shows the absolute maximum and minimum range of values derived from our model
for the obscured star formation rate density in comparison with observed optical/UV points in a
H$_{0}$=50 km s$^{-1}$ Mpc$^{-1}$ and $q_{0}$=0.5 cosmology for comparison to other
works. Data points are from 
\citet{Lil96, Mad96, Con97, Cow99, St99, Yan99} (black plus, blue diamonds, green crosses, inverted
triangles, red crosses and purple triangle respectively). 
The dotted black line is
the model of \citet{Xu00b}.
The lower plot shows our three models for the obscured SFR
with the observed UV points as dotted symbols and the
extinction corrected estimates from \citet{Mad98, Meu99, St99, Tho00} as blue diamonds, red square,
black crosses, filled red squares respectively.
Also shown is the rate derived from ISO observations of the CFRS field \citep{Flo99} as
filled red circles, and estimates from the radio and submillimeter by \citet{bar00} as
filled black squares.
Our three evolutionary models are shown as a solid line (pure luminosity), broken red line
(pure density) and broken purple line (luminosity+density). 
We assign a 1$\sigma$ uncertainty of 50\%
to our estimate of the dust enshrouded star formation rate. We emphasize that our models
place only a strong upper limit on the SFR at $z>2$ and a drop off with redshift to agree with the
extinction corrected optical/UV measurements is consistent with both the submillimeter
counts and the CIRB spectrum.
}
\label{sfrhis}
\end{figure}
\fi

\ifpp
\begin{figure}[ptb]
\centerline{\psfig{file=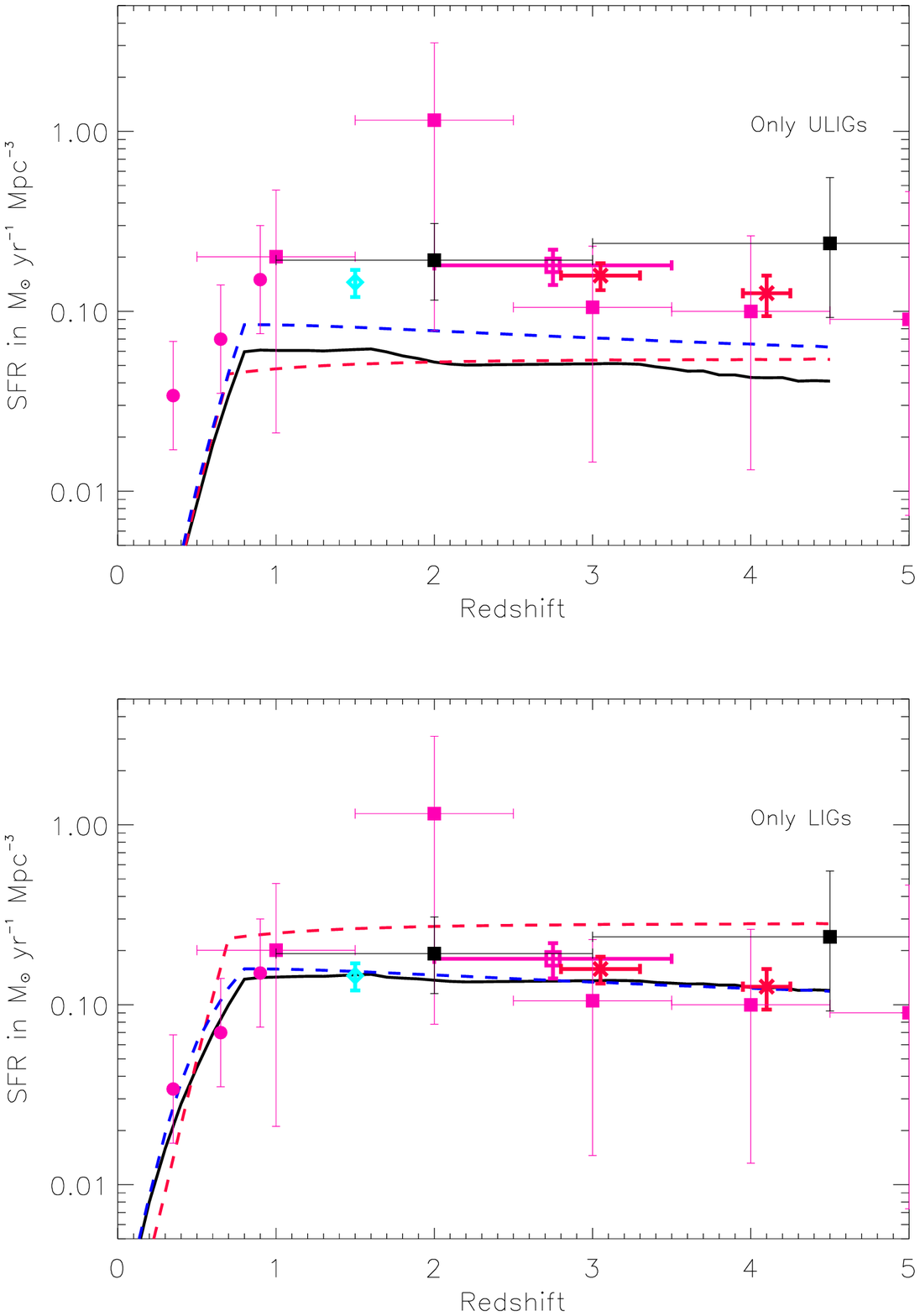,height=6.0in}}
\caption{
Plots showing the contribution to the dust enshrouded star formation rate
density from only LIGs and ULIGs as derived from our three models. A comparison
with Figure \ref{sfrhis} shows that the contribution from normal/SB galaxies with $L_{IR}<10^{11}~L_{\sun}$
is relatively small at $z>0.5$, $\sim5-30$\% depending on the model.
Symbols are the same as those in Figure \ref{sfrhis}.
}
\label{sfrhis2}
\end{figure}
\fi

Recently, \citet{Xu00b} has developed a multiparameter model
where the 25~\um\ luminosity function of \citet{Shup98} is partitioned into three
components - starburst, late-type galaxies and AGN and each component is evolved
independently of the other. Specifically, they evolved the starburst population
in luminosity as $(1+z)^{4.2}$ and in density as $(1+z)^2$ out to $z=1.5$. The late-type
galaxy population was evolved in luminosity as $(1+z)^{1.5}$ while the galaxies with AGNs
evolve in luminosity as $(1+z)^{3.5}$. Beyond $z=1.5$, all the components drop-off as $(1+z)^{-3}$.
Using a 25~\um\ to IR luminosity conversion based on IRAS data, we have converted their evolution 
for starburst and late-type galaxies into
a star formation rate and compared it with ours. This is shown as the black dotted line in the
upper plot of Figure \ref{sfrhis}. We find that their derived rates between redshifts of 1 and 2.5
are inconsistent with our models. The motivation for this peak is not clear since
only the CIRB and the SCUBA counts place constraints on the evolution at this redshift range
and both
can be reproduced very well by an almost flat evolutionary history at $z > 0.8$ (See Section 3). 
However,
their evolution at $z < 1$ agrees reasonably well with ours since both are principally constrained
by the 15~\um\ ISOCAM counts. Furthermore,
their decline in the SFR at high redshift ($z > 2$) is similar but below our lower limit.

By integrating our comoving star formation rate density over redshift and thereby cosmic time, we can
derive the density of stars and stellar remnants and 
compare it with the total baryon density in the local Universe.
Stellar lifetimes were chosen for solar metallicity stars \citep{bres93} while the
mass of remnants was chosen using the recipe of \citet[][and references therein]{pran98}.
If a Salpeter IMF is assumed, then the model predicts a local density of baryons
of about 1.0$\times10^{9}~M_{\sun}~Mpc^{-3}$ which is a factor of 2 in excess of the
value
of 5$\pm$3$\times 10^{8}~M_{\sun}~Mpc^{-3}$ estimated by \citet{fuku98}.
The model also predicts that 
100\% of the local stars and remnants would have been produced at a redshift $z<2.0$.
If we instead use the shape of the IMF
below 1 M$_{\sun}$ suggested by \citet{gou96} which reduces the number density of low
mass stars, then the
density of stars and remnants resulting from the model is 7.5$\times 10^{8}~M_{\sun}~Mpc^{-3}$, in 
agreement with the local density. \citet{Mad00} argued for a similar IMF based on their analysis of the
total extragalactic background light. 
Interestingly, our model predicts that the local baryon density in stars and remnants, 
derived by integrating
the star formation in ULIGs with a redshift distribution as shown in Figure \ref{sfrhis2},
is similar to that seen in local spheroids, suggesting that high redshift
infrared luminous galaxies may be the progenitors of present day spheroids.

\section{Conclusions}

A variety of observational data at mid-infrared through submillimeter wavelengths
trace the fraction of emission from stars that is thermally reprocessed by dust.
By using the counts of galaxies at these wavelengths, it is possible to estimate
the amount of star formation that is enshrouded by optically thick HII regions
and thereby invisible to observations at ultraviolet and visible wavelengths.
In addition, the spectrum of the cosmic infrared background at mid- and far-infrared
wavelengths places an upper limit on the fraction of starlight that has undergone
thermal reprocessing by dust.

We have developed a set of template spectral energy distribution (SEDs) for galaxies as a function
of infrared luminosity, which reproduce existing data 
at 0.44, 7, 12, 15, 25, 60, 100, and 850~\um\ from ISO, IRAS and SCUBA on nearby galaxies.
The 15~\um\ local luminosity function was then evolved with redshift, 
considering both luminosity and density evolution models into account,
and using the template SEDs to fit the observed counts
at 15~\um, 90~\um, 170~\um\ and 850~\um. A number of evolutionary models provide reasonable fits
to the data and the spectrum of the CIRB. The principal reason for this is that
all the long wavelength surveys are typically sensitive to only the most luminous galaxies
($L_{IR} > 10^{11}~L_{\sun}$) at $z>0.5$. 
So, evolutionary models that result in similar
luminosity functions at $L_{IR} > 10^{11}~L_{\sun}$ are degenerate. However, our models
accurately constrain the comoving number density of these luminous galaxies
as a function of redshift.
In the local universe, it is these galaxies, many of which show morphological signatures
of interaction, that show an infrared-determined star formation rate (SFR) that is about an 
order of magnitude higher than the corresponding UV-determined SFR.
By integrating the infrared luminosity of these luminous galaxies we then obtain an estimate
of the dust enshrouded star formation rate. The dust enshrouded star formation rate
density
appears to peak at a much lower redshift than previously thought, at $z=0.8\pm0.1$ with a value of
0.25$^{+0.12}_{-0.1} M_{\sun}~yr^{-1} Mpc^{-3}$ and remains approximately constant at least till $z\sim2$.
Any drop off at a lower redshift would result in an underestimate of the 850~\um\ galaxy counts.
Although our models do not constrain the evolution of the faint-end ($L_{IR} < 10^{10}~L_{\sun}$)
of the luminosity function, their net contribution to the high-redshift dust enshrouded
star-formation is negligible
as can be seen in the range of evolutionary models considered. The evolution at $z > 2$ is
constrained much more weakly. Having a constant SFR between redshifts of 0.8 and 4 is consistent
with the CIRB spectrum and the submillimeter counts as is a decay by a factor of 7 between
redshift 2 and $\sim$5. However, we find that there is excellent agreement between our 
luminosity function and the
infrared luminosity function derived from extinction correction to optical/UV observations
of Lyman-break galaxies at $z\sim3$. This suggests that dust obscuration is significant
even at $z > 3$ and that the dust enshrouded star formation
rate is constant to within a factor of 2 between redshifts 2 and 4.

The models also provide a census of the luminosity of galaxies that contribute to the counts at
different wavelengths, their redshift distribution and the relative contribution to the cosmic
infrared background at $\lambda > 5$~\um\ as a function of redshift. Furthermore, we find
that ultradeep
observations with SIRTF at 24~\um\ down to a sensitivity of 25~$\mu$Jy
can potentially break the degeneracy in the 
evolutionary models by
detecting galaxies with $L_{IR}\sim10^{11.5}~L_{\sun}$
out to $z\sim2.5$, well beyond the turnover redshift of 0.8 that is derived from our models.

\acknowledgements

RC wishes to thank Harland Epps and Rodger Thompson for kindly funding this 
research through NASA grant NAG5-3042. 
DE wishes to thank the American Astronomical Society for its support
through the
Chretien International Research Grant and Joel Primack and David Koo
for supporting his research through NASA grants NAG5-8218 and NAG5-3507.
We wish to acknowledge Pierre Chanial
for collating published data from a large number of surveys and making
them available to us.

\end{document}